\documentclass[12pt]{iopart}

\expandafter\let\csname equation*\endcsname\relax
\expandafter\let\csname endequation*\endcsname\relax
\usepackage{amsmath}

\usepackage{array}
\usepackage{booktabs, longtable}
\usepackage{hyperref}
\usepackage{xcolor}
\usepackage{lscape}
\usepackage{multirow}
\usepackage{graphicx}
\usepackage{caption}
\usepackage{subcaption}

\usepackage[capitalize]{cleveref}

\usepackage{etoolbox}
\newtoggle{anonymous}
\togglefalse{anonymous}

\newcommand{\cwV}{3.2cm}
\newcommand{\cwS}{1.7cm}
\newcommand{\cwT}{2.5cm}
\newcommand{\cwE}{3.3cm}
\newcommand{\cwP}{3.2cm}
\newcolumntype{V}{>{\raggedright}p{\cwV}}
\newcolumntype{S}{>{\raggedright}p{\cwS}}
\newcolumntype{T}{>{\raggedright}p{\cwT}}
\newcolumntype{E}{>{\raggedright}p{\cwE}}
\newcolumntype{P}{>{\raggedright}p{\cwP}}

\newcommand{\appFootnote}{\footnotesize{***, **, and * represent statistical signiﬁcance at \(\textit{p} < 0.001\), \(\textit{p} < 0.01\), and \(\textit{p} < 0.05\), respectively.}}

\begin{document}

\title[]{Social and Economic Impact Analysis of Solar Mini-Grids in Rural Africa: A Cohort Study from Kenya and Nigeria}

\iftoggle{anonymous}{}{
	\author{A.T. Carabajal$^1$, A. Orsot$^2$, M.P.E. Moudio$^3$, T. Haggai$^4$, C.J. Okonkwo$^5$, G.T. Jarrard III$^1$, and N.S. Selby$^2$}
	\address{$^1$Renewvia Energy Corporation, 2951 Flowers Rd. S., Suite 217, Atlanta, GA 30341}
	\address{$^2$Renewvia Solar Ethiopia PLC, Addis Ababa, Ethiopia, Bole sub-city, Woreda 03, House No. 174/175, Saya Building, 5th Floor}
	\address{$^3$Department of Industrial Engineering and Operations Research, UC Berkeley, 2521 Hearst Ave, Berkeley, CA, USA, 94709}
	\address{$^4$Department of Anthropology, Gender and African Studies, University of Nairobi, PR98+JC9 University Way, Nairobi}
	\address{$^5$Planetary Health Management Mission Pathway, Entrepreneurial Leadership Program, Africa Leadership University Rwanda Campus,
	Bumbogo, Kigali, Innovation City, Kigali, Rwanda}
	\ead{\mailto{nicholas.selby@renewvia.com}}
}
\vspace{10pt}
\begin{indented}
\item[]December 2023
\end{indented}

\begin{abstract}
This study presents the first comprehensive analysis of the social and economic effects of solar mini-grids in rural African settings, specifically in Kenya and Nigeria. A group of 2,658 household heads and business owners connected to mini-grids over the last five years were interviewed both before and one year after their connection. These interviews focused on changes in gender equality, productivity, health, safety, and economic activity. The results show notable improvements in all areas. Economic activities and productivity increased significantly among the connected households and businesses. The median income of rural Kenyan community members quadrupled. Gender equality also improved, with women gaining more opportunities in decision making and business. Health and safety enhancements were linked to reduced use of hazardous energy sources like kerosene lamps. The introduction of solar mini-grids not only transformed the energy landscape but also led to broad socioeconomic benefits in these rural areas. The research highlights the substantial impact of decentralized renewable energy on the social and economic development of rural African communities. Its findings are crucial for policymakers, development agencies, and stakeholders focused on promoting sustainable energy and development in Africa.
\end{abstract}

\vspace{2pc}

%
\submitto{Environmental Research: Infrastructure and Sustainability}
%
\maketitle
%
%


\section{Introduction}
\label{sec:intro}
\begin{figure}[th]
	\centering
	\includegraphics[width=0.8\textwidth]{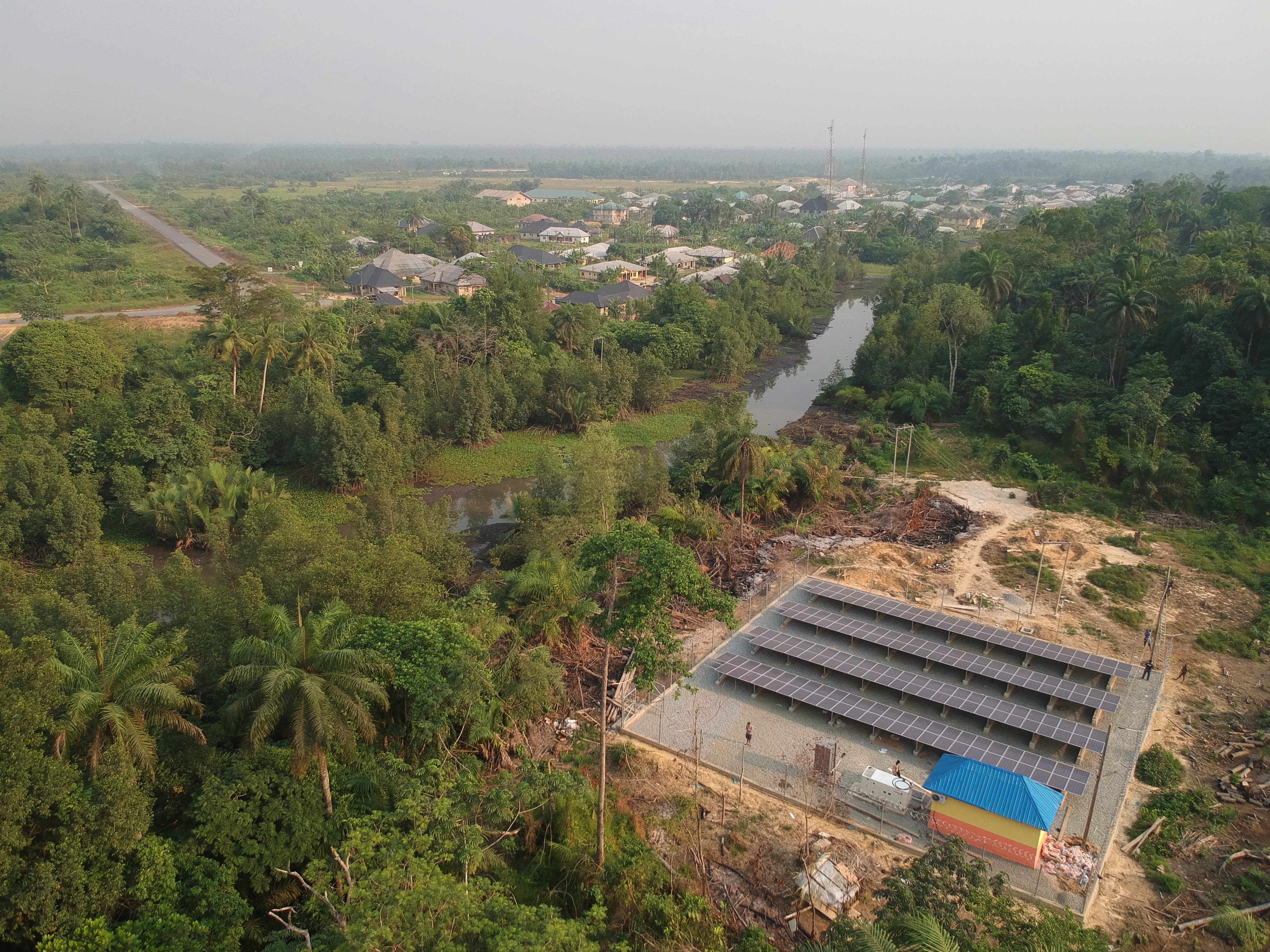}
	\caption{A solar mini-grid in Bayelsa, Nigeria}
	\label{fig:mini-grid}
\end{figure}

Sub-Saharan Africa (SSA) has vast renewable energy resources, including solar, wind, hydropower, and geothermal \cite{hafner2018prospects}. However, these resources remain largely untapped, and the region is still heavily reliant on fossil fuels due to factors including a lack of infrastructure, financing, and technical expertise \cite{irena2022renewable}. Decentralized renewable energy solutions, especially renewable mini-grid (RMG) systems, can offer a promising alternative to grid-based electricity in SSA \cite{moro2018ensuring, rasagam2018delivering}. RMG systems like the one illustrated in \cref{fig:mini-grid} are typically smaller in scale and can be installed in rural areas where grid extension is not feasible or cost-effective. RMG systems can also be operated by local communities, empowering these communities and promoting economic development \cite{deshmukh2009role, edwards2018role}.
Several studies have investigated the potential of RMGs as decentralized renewable energy solutions in SSA \cite{rabetanetiarimanana2018pv, okunlola2018assessment, moner2018electrification}. These studies have found that RMGs can have a significant positive impact on access to electricity, livelihoods, and economic development \cite{dawoud2018hybrid}. RMGs are small power grids that typically serve a few hundred to a few thousand households and businesses. Often powered by renewable energy sources, such as solar or wind, they can provide a reliable and affordable source of electricity to communities that are not connected to the national grid.

Rural electrification is the process of providing fast, reliable, and affordable access to electricity to residents in rural areas who currently lack it. Access to electricity is seen as key to reducing poverty and is a critical component of economic development and poverty alleviation due to its ability to improve access to and quality of essential services such as education, healthcare, and clean water \cite{hussein2012analysis, hanjra2009reducing}. It can also create new employment opportunities and boost agricultural productivity. SSA has the lowest electrification rate in the world and the majority of those without access live in rural areas due to lack of financial ability to extend the grid, population density, and other social and cultural factors \cite{zebra2021review}. Data from a 2019 World Bank report shows that 14 of the 20 countries in the world with the highest deficit in electricity connections are African countries \cite{Tracking_SDG7_Report_2019}. Only $45\%$ of the population in SSA currently have access to electricity. The electrification coverage is expected to rise to only about $60\%$ of the SSA population by 2030 given current conditions \cite{valickova2021costs}. Despite representing approximately $14\%$ of the world's population, SSA accounts for only about $4\%$ of global energy consumption \cite{falk2021socio}.

Rural electrification can significantly benefit communities \cite{chakravorty2016lighting}. According to \cite{falk2021socio}, the introduction of electrification via RMG has had significant positive impacts on socioeconomic factors such as health care and economic development. RMG systems promote faster and more flexible energy supply, especially when integrated with additional components such as electricity storage in a hybrid system \cite{mazzeo2021literature}. The slow extension of conventional grid systems has led to increased interest in RMGs, a decentralized alternative not connected to the public grid and generating electricity based on various technologies \cite{korkovelos2020retrospective}. According to \cite{valickova2021costs}, the incremental cost of providing electricity access to households is below the cost of alternatives such as kerosene lighting.

Various socioeconomic factors affect rural electrification in SSA. According to \cite{poblete2021model}, income levels are a key driver of rural electrification. As incomes rise, households are more likely to be able to afford more electricity and appliances. The rising demand in RMGs due to growing populations, industrial growth, and government policies can play a significant role in promoting rural electrification \cite{antonanzas2021state}. Technological advancements also play an important role in rendering the electrification process more feasible and affordable to provide electricity to rural areas. Moreover, digital and information technologies have several positive effects on the development and regulation of efficient energy consumption \cite{shabalov2021influence, Consultant_Klooss_Consultant_2021}.

While existing research has acknowledged the significance of social impacts stemming from RMG implementation in SSA, there remains a notable absence in empirical measurements \cite{eales2018social}. The prevailing studies predominantly focus on technical and economic aspects, neglecting the quantification of social ramifications \cite{pueyo2018impact, wassie2021socio, uwineza2021analysis, lee2020experimental}. A recent study in the Gbamu Gbamu village in Nigeria uses statistical tests to measure the financial impact of RMGs on local businesses using predominantly economic factors including gender, marital status, household size, age, education level, years of business establishment, hours of operation, building tenure, capital source, number of employees, generator ownership, and the days of operation \cite{babalola2022socio}. 

A few studies have attempted to formally quantify social impacts of RMGs on communities. In \cite{liu2021enabling}, researchers apply a mediation model to illustrate how community engagement results in a $1\%$ increase in the perceived renewable energy potential, leading to a $0.195\%$ increase in perceived poverty reduction. These results suggest that community empowerment is indispensable in creating electricity demand and delivering development impact of renewable RMGs in the context of deep poverty. In another study carried out in Kyenjojo District in western Uganda, the authors incorporated social factors in a study of RMG  operation, causes of failure, sources of discomfort to customers, and customer behavior \cite{cartland2022socio}. Moreover, researchers in \cite{duran2021analysis} conducted a systematic review of diverse RMG projects aimed at extracting qualitative insights into the factors driving project success and community benefits. Subsequently, these findings were empirically validated, enabling the identification of key factors contributing to the success and cost-effectiveness of RMG projects \cite{nyarko2023drivers}. These studies have highlighted the necessity for broader assessments that go beyond solely examining economic impacts and RMG operational challenges in rural areas relying on RMGs. We aim to broaden this perspective by considering both economic and social factors like gender equality and community residents' safety. This shift recognizes the interconnection between socioeconomic elements, addressing a significant gap in understanding the overall transformational impact of RMG initiatives within rural settings.

Our study aims to bridge the gap in empirical measurement of social impact of RMGs on rural communities by introducing a comprehensive framework that delineates five crucial axes encompassing gender equality, productivity, health, safety, and economic activity. By leveraging a survey designed specifically around these axes, we endeavor to quantify the multifaceted impacts of solar RMGs on various social and economic factors within rural African communities. Unlike previous single-region analyses, our research embraces a comparative approach, drawing insights from two distinct countries, Kenya and Nigeria. This deliberate expansion of our dataset yields a richer understanding of the nuanced effects of RMG interventions across diverse sociocultural landscapes. The rest of the paper is organized as follows. In \cref{sec:methodology}, we discuss the model, data, and estimation techniques. The results and findings are presented in \cref{sec:results}. Further discussion and methodology limitations are shown in \cref{sec:discussion}. We conclude in \cref{sec:conclusion}.
\section{Methodology}
\label{sec:methodology}
\subsection{Project Description and Data Collection}
Between 2021 and 2023, we installed solar mini-grids in and conducted a study across 22 communities in Nigeria and Kenya to assess if there were any significant changes regarding the quality of life of the populations connected to their mini-grids. The capacity of the mini-grids varied, starting from 6.24 kWp with 14.8 kWh of battery storage, up to 541 kWp accompanied by 1.10 MWh of battery storage. To collect the necessary information, we used surveys composed of semi-structured questionnaires that captured quantitative and qualitative data such as demographics, access to education for children, access to clean drinking water, creation of jobs and businesses, and economic opportunities for women, as well as many others. Primary data was collected from a variety of respondents categorized either as households or commercial and institutional organizations (e.g. businesses, schools, clinics, etc.) through face-to-face interviews by independent, third-party field enumerators.

This study examined the impact of the mini-grids on the key performance indicators (KPIs) by comparing responses before and one year after installing a solar mini-grid in the respective communities. Due to the lack of data from communities without a solar mini-grid, only pre-and post-treatment data have been used in this study. The pre-treatment survey was conducted between 2021 and 2022, while the post-treatment survey was conducted between 2022 and 2023.

\subsection{Survey Design} 
This study aimed to investigate the impact of installing mini-grids in rural communities in Sub-Saharan Africa, focusing on five KPIs: gender equality, productivity, safety, health, and economic growth.

A cohort study design was employed, enabling an analysis of the mini-grid's effects on individuals over time. This approach was particularly chosen for its ability to observe changes between pre- and post-mini-grid installation. The survey targeted both commercial and residential mini-grid users in communities shortly before connection and one year after connection to new mini-grids.

The survey was developed based on literature review and global KPIs relevant to mini-grid stakeholders. No pilot testing was conducted. A diverse set of question types was utilized, including demographic queries, Likert-scale items, interval and quantity-specific questions, and open-ended questions. This mix aimed to capture both quantitative and qualitative aspects of the mini-grid's impact. To ensure reliability and validity, the survey employed clear, concise, and neutrally-worded questions. Professional survey administrators were engaged to maintain consistency and integrity in data collection.

Surveys were carried out in person by independent, third-party surveyors who used CommCare electronic forms to minimize transcription bias. Participants were briefed and consent obtained through signed forms, ensuring ethical compliance. No incentives were offered.

Data analysis involved various statistical tests, including paired $t$-tests, Wilcoxon signed-rank tests, and linear regression, among others, using R software. This comprehensive approach aimed to rigorously assess the mini-grid's impact across multiple dimensions.

\subsection{Analysis Techniques}
\subsubsection{Paired Samples \textit{t}-Test}
\hfill \break
The paired samples $t$-test is a parametric statistical method to determine whether the mean difference of paired measurements is 0 or not. It follows the assumptions that the observations are independent, the paired differences are approximately normally distributed, and there are no extreme outliers in the differences. The paired samples $t$-test includes the following null and alternative hypotheses:
\begin{equation}
\begin{array}{c}
	H_0 : \mu_d = 0 \\
	H_1 :  \mu_d \neq 0
\end{array}
\end{equation}
where $\mu_d$ is the mean of differences from the pairs, $H_0$ the null hypothesis stating the mean paired difference is equal to 0, and $H_1$ the alternative hypothesis stating that the mean paired difference does not equal 0.

The paired samples $t$-test utilizes the $t$ statistic
\begin{equation}
	t = \frac{\bar{d}\sqrt{n}}{\sigma_d}
\end{equation}
where $\bar{d}$ is the mean value of the differences between paired samples, $\sigma_d$ the standard deviation of the differences between paired samples, and $n$ is the sample size.

\subsubsection{Wilcoxon Signed-Rank Test}
\hfill \break
The Wilcoxon signed-rank test is a statistical method to compare two dependent samples from paired data. While it assumes the distribution of the differences is symmetric, it does not assume any specific distribution of the samples themselves and serves as a non-parametric equivalent to the paired samples $t$-test, particularly applicable to categorical variables with meaningful differences between ranks (i.e. ordinal data). The test is evaluated taking into account both the sign and magnitudes of observed differences. The Wilcoxon signed-rank test is implemented using the null and alternative hypotheses:
\begin{equation}
\begin{array}{c}
	H_0: \mathrm{M} = 0   \\
	H_1: \mathrm{M} \neq 0
\end{array}
\end{equation}
where M is the median of the paired differences, $H_0$ the null hypothesis stating no difference between paired observations, and $H_1$ the alternative hypothesis stating a significant difference between paired observations.

The Wilcoxon signed-rank test utilizes the \textit{W} statistic
\begin{equation}
	 W = \mathrm{min}(T_-, T_+)
\end{equation}
where $T_-$ is the sum of the negative differences and $T_+$ is the sum of the positive differences.

\subsubsection{Sign Test}
\hfill \break
The Sign test is a non-parametric statistical method designed to determine if two dependent samples, ordered in pairs, are of equal magnitudes. Unlike the Wilcoxon signed-rank test, it does not assume symmetry and considers only the direction of change, making it suitable for categorical variables where arithmetic differences are not meaningful. It is often viewed as a less powerful test since it does not measure the magnitude of differences between pairs, but it is still useful for assessing the significance of observed changes. Although the Wilcoxon signed-rank test and the Sign test operate similarly, the choice between them depends on the data characteristics. The Wilcoxon signed-rank test is preferable for differences that are approximately normally distributed and have meaningful magnitudes, while the Sign test is more appropriate in other cases. The Sign test is implemented using the null and alternative hypotheses:
\begin{equation}
\begin{array}{rl}
	H_0 : & \mathrm{The\ signs\ of\ +\ and\ -\ of\ differences\ are\ of\ equal\ size} \\
	H_1 : & \mathrm{The\ signs\ of\ +\ and\ -\ of\ differences\ are\ not\ of\ equal\ size}
\end{array}
\end{equation}

The sign test utilizes the test statistic
\begin{equation}
	Z = \frac{(2S - n)\sqrt{n}}{n}
\end{equation}
where $n$ is the total number of signs, ignoring 0s, and $S$ the number of less frequent signs.

\subsubsection{McNemar's Test}
\hfill \break
McNemar's test is a non-parametric test used to analyze paired nominal data. It is a test on a $2\times 2$ contingency table that checks the marginal homogeneity of two dichotomous variables. The test requires one nominal variable with two categories and one independent variable with two dependent groups.

\begin{table}[ht]
\centering
\begin{tabular}{ |c|c|c|c| } 
	\hline
	 & Post: Yes & Post: No & Total \\
	\hline
	Pre: Yes & a & b & a+b \\ 
	Pre: No & c & d & c+d \\
	Total & a+c & b+d & n \\
	\hline
\end{tabular}
\caption{McNemar Contingency Table}
\label{tab:mcnemar}
\end{table}

We use \cref{tab:mcnemar} to calculate the $\chi^2$ goodness-of-fit statistic with the following null and alternative hypotheses:
\begin{equation}
\begin{array}{c}
	H_0: P_b = P_c   \\
	H_1: P_b \neq P_c  
\end{array}
\end{equation}
where \(H_0\) is the null hypothesis stating that the two marginal probabilities, $P_b$ and $P_c$, for each outcome are the same, and \(H_1\) is the alternative stating otherwise.

McNemar's test utilizes the $\chi^2$ statistic:
\begin{equation}
	\chi^2 = \frac{(b-c)^2}{(b + c)}
\end{equation}

\subsubsection{Pearson Correlation Coefficient}
\hfill \break
The Pearson correlation coefficient is a measure representing the strength of association between two variables and the direction of the relationship. It produces a value between -1 and +1, serving as a descriptive statistic. A value nearing +1 suggests that a change in one variable will lead to a similar directional change in the other, whereas a value approaching -1 indicates that altering one variable results in a change in the opposite direction for the other. We obtain the Pearson correlation coefficient $r$ using the formula:
\begin{equation}
	 r = \frac{\sum((x_i - \Bar{x})(y_i - \Bar{y}))}{\sqrt{\sum(x_i - \Bar{x})^2\sum(y_i - \Bar{y})^2}}
\end{equation}
where $x_i$ is the value of the $i^\mathrm{th}$ predictor in a sample, $\bar{x}$ is the mean of the values of the predictor variable, $y_i$ is the value of the $i^\mathrm{th}$ response in a sample, and $\bar{y}$ the mean of the values of the response variable.

\subsubsection{Linear Regression \textit{t}-Test}
\hfill \break 
The simple linear regression is a statistical method to evaluate the relationship between a predictor variable and a response variable by finding the equation of a ``best-fit'' line that minimizes the sum of squared residuals between it and the data. It estimates the nature of the relationship, either positive or negative, and the expected change in the response based on a change in the predictor. A one-sample $t$-test is applied to the slope to determine if said relationship is statistically significant given the following null and alternative hypotheses:
\begin{equation}
\begin{array}{c}
	H_0: \beta_1 = 0 \\
	H_1: \beta_1 \neq 0
\end{array}
\end{equation}
where \(H_0\) is the null hypothesis stating that there is no relationship between outcome and predictor, \(H_1\) is the alternative stating otherwise, and $\beta_1$ is the slope of the ``best-fit'' line given by
\begin{equation}
	\hat{y} = \beta_0 + \beta_1x
\end{equation}
where \(\hat{y}\) is the expected value of response; \(\beta_0\) is the intercept, i.e. the expected value of response when the predictor is 0; \(\beta_1\) is the slope coefficient, i.e. the average change in the response given a unit increase in the predictor; and $x$ is the value of predictor.

The linear regression $t$-test utilizes the $t$-statistic
\begin{equation}
	t=\frac{r\sqrt{n-2}}{\sqrt{1-r^2}}
\end{equation}
where $r$ is the Pearson correlation coefficient and $n$ is the number of data points $(x,y)$. Note that this $t$-statistic has a $t$-distribution with $n-2$ degrees of freedom if the null hypothesis is true.

In this study, regression analysis was implemented at two levels:
\begin{itemize}
	\item At the individual customer level, investigating the relationship between average monthly electricity consumption for a specific user, the predictor, and various survey question responses; and
	\item At the community level, investigating the relationship between predictor variables of mini-grid PV system capacity, total number of customers in the community, and total mini-grid capital expenditure (CAPEX), and various aggregated survey question responses. For the purposes of this paper, ``mini-grid PV system capacity,'' or simply ``mini-grid capacity'' or ``PV Size'' represents the maximum amount of power that the solar panels can produce under ideal conditions (STC), usually given in units of kilowatt-peak (kWp). For example, a 10-kWp solar mini-grid would be expected to produce up to 10 kilowatts of power during peak sunlight conditions.
\end{itemize}

For ordinal variables with three levels such as -1, 0, and 1, additional transformations were carried out to determine a pertinent response variable. First, the proportion for the level of interest (e.g., ``1'' to denote an increase in schooling for girls) was obtained for each site and then multiplied by the total number of customers for the relevant customer type (e.g., households, schools, businesses, etc.) depending on the response variable.

\subsubsection{Likelihood-Ratio Test}
\hfill \break
The likelihood-ratio test is a statistical method used to assess the significance of a predictor variable in the context of logistic regression, which models the relationship between a binary response variable and a ratio predictor variable. Logistic regression predicts the log odds of the occurrence of an event by fitting data to a logistic curve. This study considers only the case of simple binary logistic regressions in which the data are fitted in a probabilistic sense to a function of the form:
\begin{equation}
	p(x)=\frac{1}{1+\mathrm{exp}(-t)}
\end{equation}

The likelihood-ratio test compares the goodness-of-fit of two models: one ``full model'' that includes the predictor variable (i.e. $t=\beta_0+\beta_1x$) and one ``reduced model'' that does not (i.e. $t=\beta_0$). The test evaluates whether the inclusion of the predictor significantly improves the model. The null hypothesis for this test is that the predictor variable has no effect, and the reduced model is sufficient. The alternative hypothesis for this test is that the predictor variable has a significant effect, and the full model is more appropriate.

In logistic regression, the likelihood of observing the given data is maximized, and the test statistic is calculated as:
\begin{equation}
	D = -2 \ln\left(\frac{\text{Likelihood of reduced model given the data}}{\text{Likelihood of full model given the data}}\right)
\end{equation}

This test statistic follows approximately a $\chi^2$ distribution with degrees of freedom equal to the difference in the number of parameters between the full and reduced models. The decision about the significance of the predictor variable is made based on the $p$-value obtained from this $\chi^2$ distribution.

In this study, we utilize the likelihood-ratio test to determine whether or not the inclusion of the average monthly electricity consumed by a customer significantly improves a logistic regression model's ability to predict binary or dichotomous survey response variables.
\section{Results}
\label{sec:results}
In this section, we highlight the most notable descriptive statistics and statistically significant findings. A detailed summary of the outcomes for each quantitative, comparative survey question is provided in the appendix.

In instances where survey questions covered multiple key performance indicators (KPIs), the findings are presented across multiple subsections. This approach ensures that each KPI is thoroughly addressed and the results are clearly communicated in their respective areas of relevance.

For the paired testing, which compared pre-connection data to post-connection data in that order, a negative result indicates a decrease over time while a positive result signifies an increase. 

In this section, where applicable, 95\% confidence intervals for a result are denoted using the "$\pm$" symbol (e.g., $10\pm1$).

\subsection{Survey Distribution}
\begin{figure}[th]
	\centering
	\includegraphics[width=0.8\textwidth]{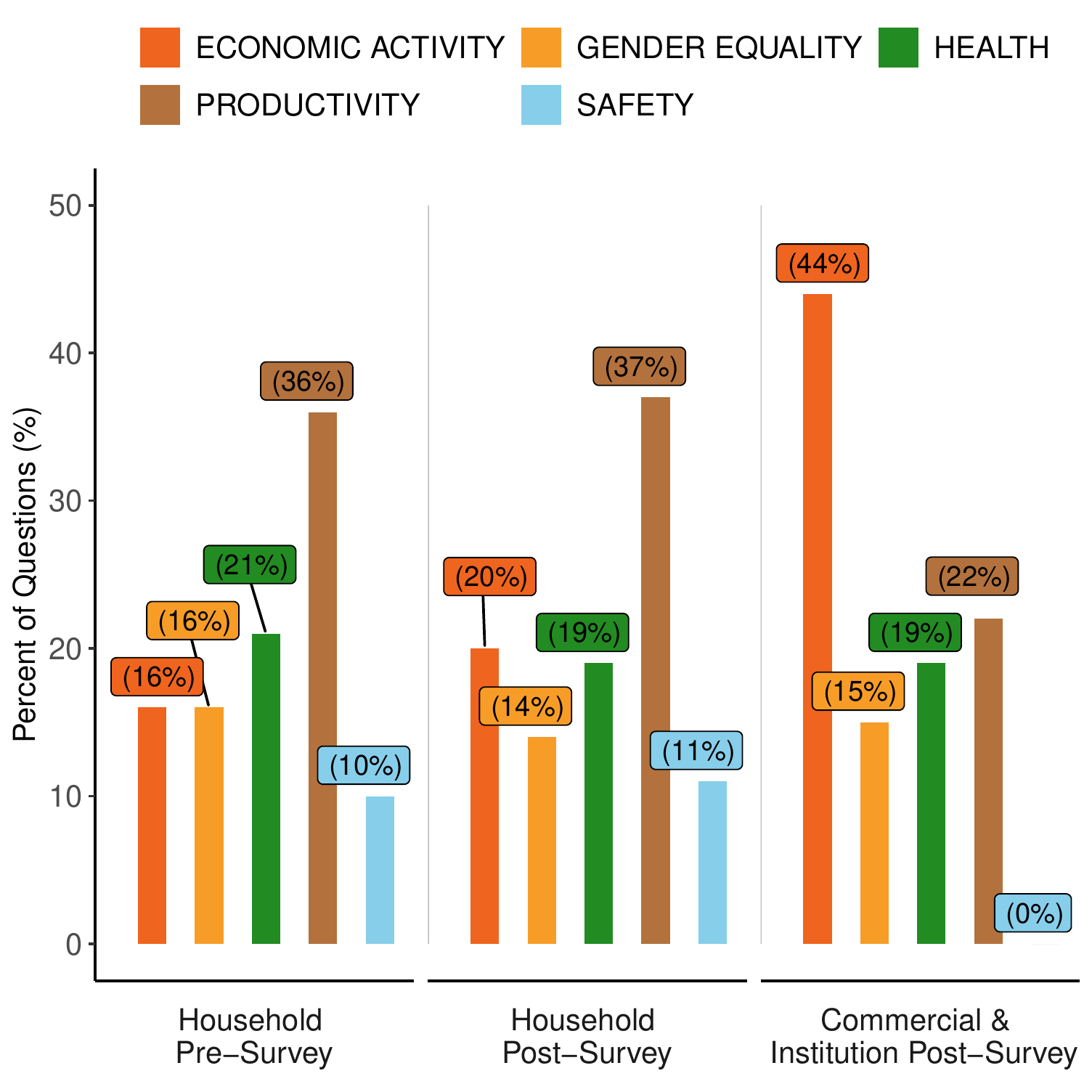}
	\caption{Distribution of Survey Questions}
	\label{fig:distrbution_questions}
\end{figure}

There are a total of 86 questions for the pre-connection household survey and 93 and 37 for the post-connection household and organizational surveys, respectively. The pre- and post-connection household surveys shared 64 questions, enabling direct comparison of individual customer responses before and after connection. The questions were distributed among the five KPIs according to \cref{fig:distrbution_questions}. Correlation tables of the responses are presented in \cref{fig:corr-pre,,fig:corr-post-hh,,fig:corr-post-ci} in the appendix.

\subsection{Characteristics of the Respondents}
Out of the 3,952 total responses from the initial survey, 564 distinct households (14.3\%) were retained for analysis. To isolate the sample from pre-connection households, the remainder of observations at post-connection and duplicate responses were excluded. At post-connection, 2,202 responses were retained from the 2,603 (84.6\%) total responses collected after a similar data validation process. From the pre and post samples, 468 respondents could be paired. The high turnover rate is mainly due to the nomadic or otherwise transitory nature of the communities surveyed.

\begin{table}[th]
\centering
	\begin{tabular}{*5c} 
		\toprule
		Variables &  & Statistic  &\multicolumn{2}{c}{Total}\\
		& &  & Pre-connection & Post-connection\\
		
		\midrule
		Sample Households  & & Count & 564 & 2202 \\
		Gender  & Male & Count & 400 & 1518\\
				& Female & Count & 164 & 678\\
				& Unidentified & Count & - & 6\\
		Age  &  & Median & 41 & 38\\
		Household Size  &  & Median & 6 & - \\
		Employment & Seasonal & Count & 360 & - \\
				   & Regular & Count & 139 & - \\
				   & Unemployed & Count & 60 & - \\
				   & Unidentified & Count & 5 & -\\
				   
		\bottomrule
	\end{tabular}
\caption{Descriptive Statistics of Households}
\label{tab:desc-stats-hh}
\end{table}

A demographic overview of the household respondents is presented in \cref{tab:desc-stats-hh}. Heads of household were responsible for responding to the survey, and the distribution based on gender shows a predominance of male respondents, with men making up 71\% of the pre-connection sample and 69\% of post-connection sample. The median age of the respondents was 41 and 38 years old for the pre- and post-survey, respectively. Before connection to the mini-grid, the median household had six members, and the breakdown of respondents' employment types was as follows: 64\% were seasonally employed, 25\% had regular employment, and 11\% were unemployed.

\begin{table}[th]
\centering
\begin{tabular}{*4c} 
	\toprule
	Variables &  & Statistic &  Post-Connection\\
	\midrule                 
	Sample Organizations  & & Count & 465 \\
	Status & In Operation & Count & 449 \\
		   & Closed & Count & 7 \\
		   & Unidentified & Count & 9 \\
	Organization Type  & Business & Count & 340 \\
					   & School & Count & 36 \\
					   & Clinic & Count & 22 \\
					   & Religious and Institution & Count & 67 \\
	\bottomrule
	\end{tabular}
\caption{Descriptive Statistics of Commercial and Institutional Customers}
\label{tab:desc-stats-com}
\end{table}

Out of the 470 total commercial and institutional responses from the initial survey, 465 distinct entities (99\%) were retained for further analysis. The remainder of duplicates and unidentifiable respondents were excluded. An overview of the socio-economic status of commercial and institutional customers is presented in \cref{tab:desc-stats-com}. The distribution based on organization type was broken down with 340 (73\%) businesses, 36 (8\%) schools, 22 (5\%) clinics, and 67 (14\%) religious institutions.

\subsection{Gender Equality}
\begin{figure}[b]
	\centering
	\begin{subfigure}[t]{0.48\textwidth}
		\centering
		\includegraphics[width=\textwidth]{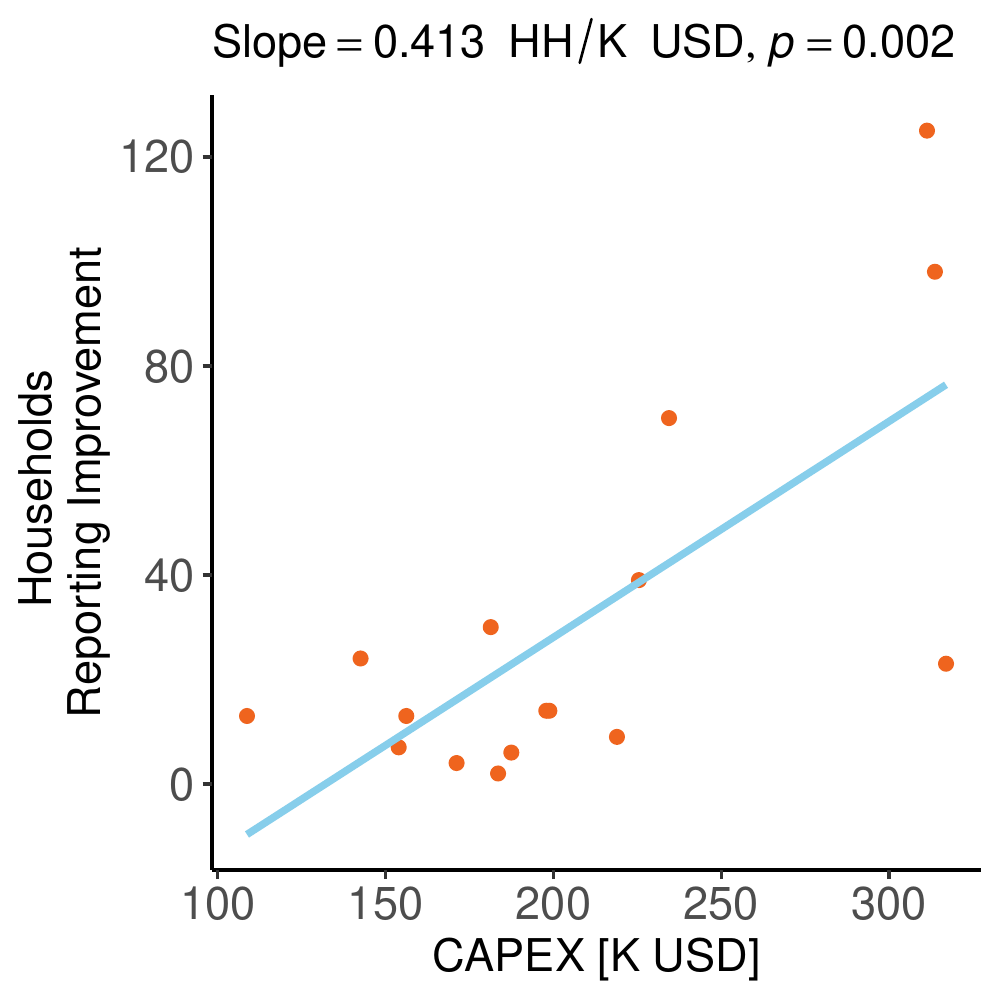}
		\caption{An additional \$50,000 CAPEX investment correlated with an additional 20 households reporting an increase in the number of girls attending school.}
		\label{fig:girls-school}
	\end{subfigure}
	\hfill
	\begin{subfigure}[t]{0.48\textwidth}
		\centering
		\includegraphics[width=\textwidth]{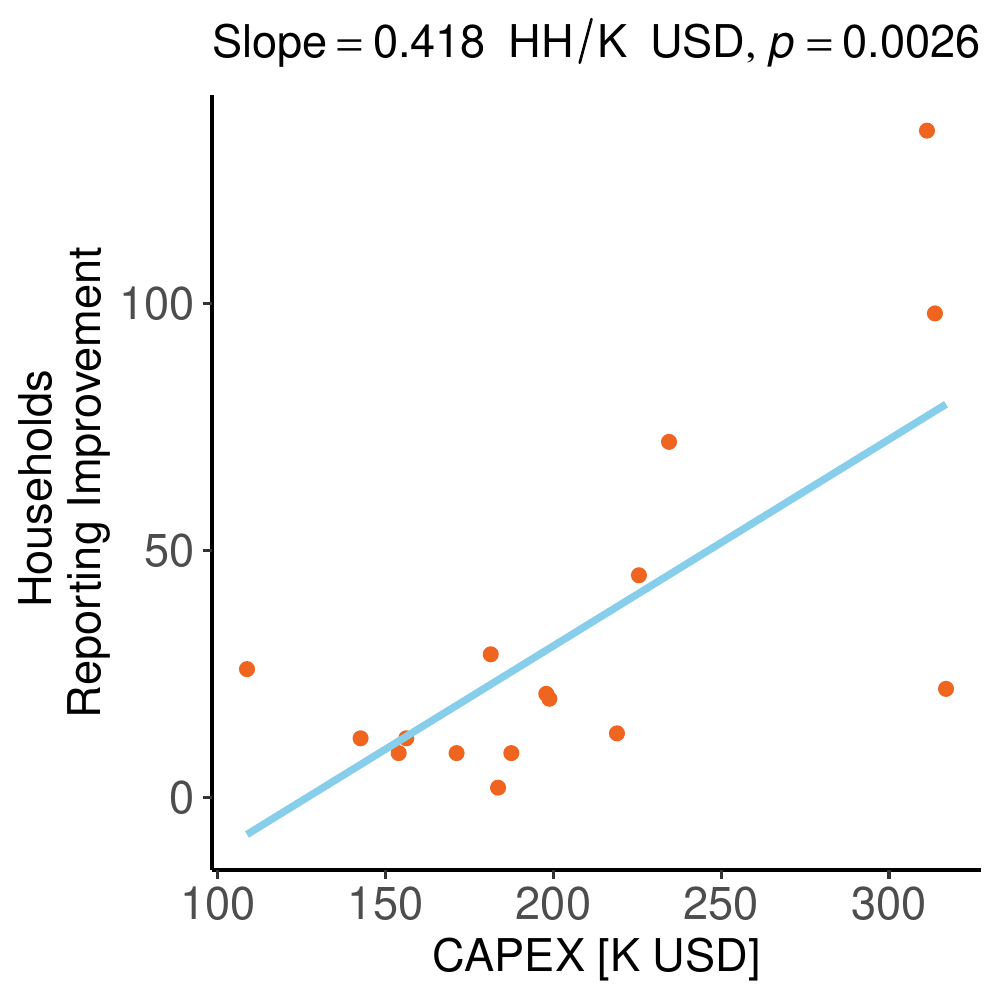}
		\caption{An additional \$50,000 CAPEX investment correlated with an additional 21 households reporting an increase in the number of boys attending school.}
		\label{fig:boys-school}
	\end{subfigure}
	\caption{Investing more money to build a larger mini-grid leads to more girls and boys attending school.}
	\label{fig:regression-schooling}
\end{figure}
The full results of the statistical analysis for questions pertaining to gender equality are available in \cref{tab:app:gender} in the appendix.

In the pre-survey, respondents reported that 793 out of 1,190 boys were attending school while 719 out of 1,080 girls were attending school, thus placing the school's enrollment rate at 67\%. After the mini-grid, 16\% and 18\% of respondents noted a positive change in the schooling for girls and boys, respectively. Initially, 144 respondents reported not being able to enroll girls into school and 141 reported similarly for the boys. Investigating the remaining barriers to parents who still did not enroll their children in school even after connection to the mini-grid, 37\% of respondents reported that the reason for keeping their children out of school was insufficient funds to support tuition and other school fees. 

Community-level regression tests on gender equality showed statistical significance. For example, as illustrated in \cref{fig:regression-schooling}, an increase of \$50,000 in total capital expenditure (CAPEX) on the mini-grid correlated with an increase of $20\pm12$ households reporting an increase in the number of girls attending school and $21\pm12$ households reporting an increase in the number of boys attending school.

\begin{figure}[t]
	\centering
	\begin{subfigure}[t]{0.48\textwidth}
		\centering
		\includegraphics[width=\textwidth]{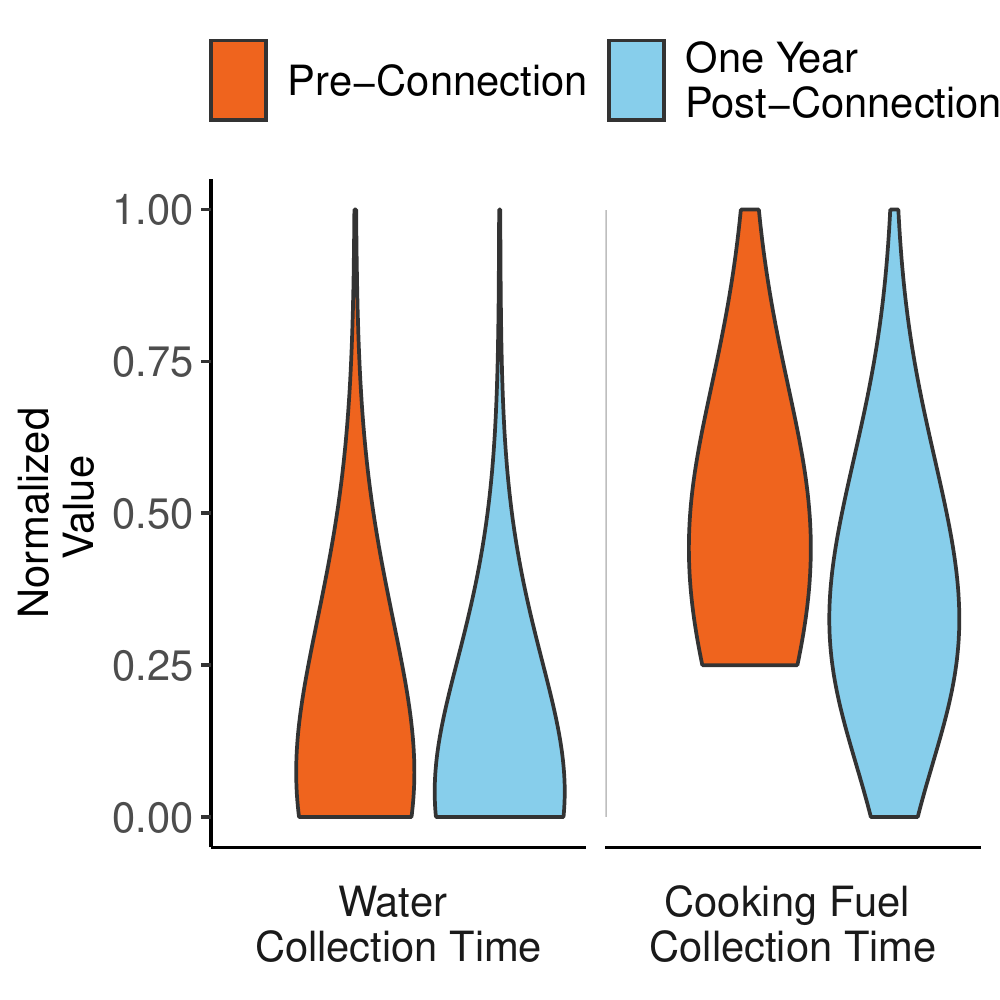}
		\caption{Residents of newly electrified communities observed statistically significant declines in water and cooking fuel collection times.}
		\label{fig:gender_equality_paired_results}
	\end{subfigure}
	\hfill
	\begin{subfigure}[t]{0.48\textwidth}
		\centering
		\includegraphics[width=\textwidth]{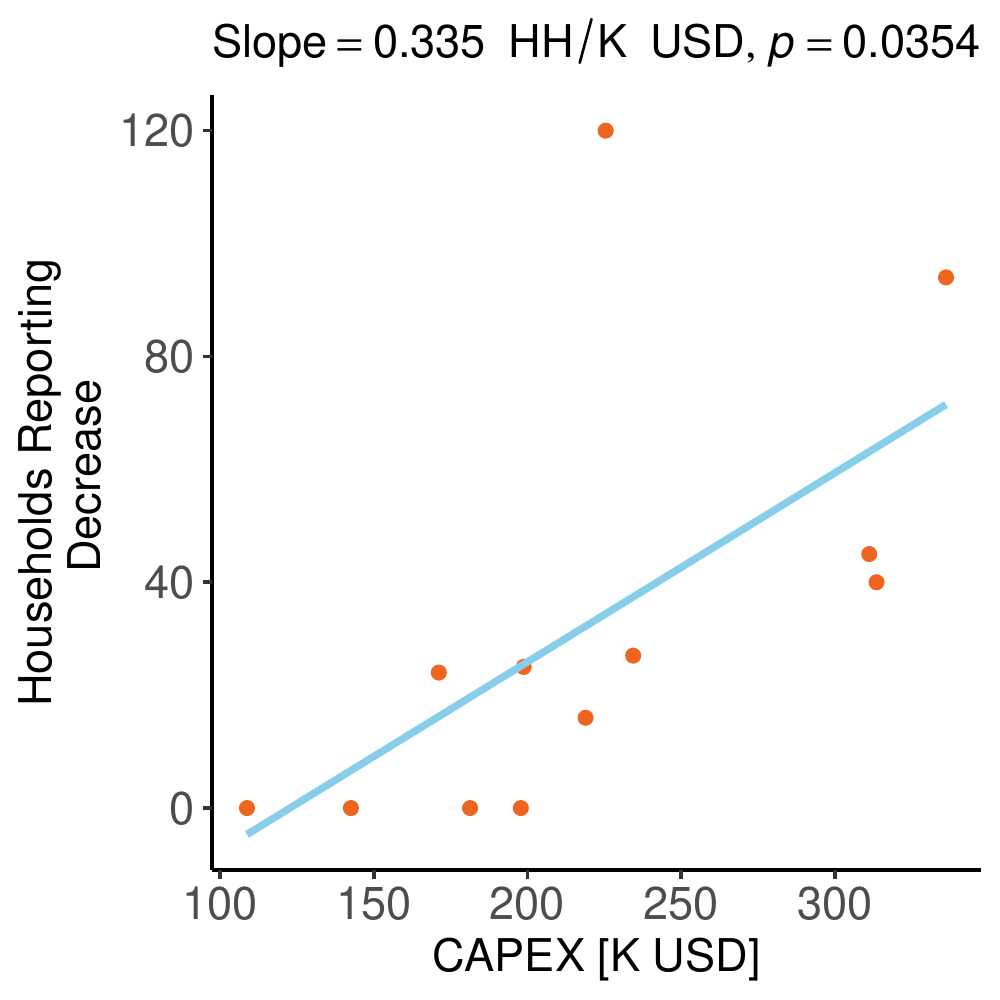}
		\caption{Increased CAPEX investment in mini-grids is correlated with ever-greater reductions in time spent collecting water.}
		\label{fig:water_collection}
	\end{subfigure}
	\caption{The presence of a mini-grid reduces the time residents have to spend on household chores.}
	\label{fig:water-and-fuel-collection}
\end{figure}

One reason behind the increase in school enrollment after connection to the mini-grid is that, in the pre-survey, 53.4\% of households reported delegating the responsibility of water collection to school-aged children. After connection to the mini-grid, only 15\% reported a similar answer in the post-survey. In the case of paired respondents, a comparable trend was noted, evidenced by a 29 percentage-point decrease in the proportion of households delegating water-fetching responsibilities to school-aged children. Overall, there was a $29\%\pm8\%$ decrease in the likelihood of school-aged children being tasked with this chore following the mini-grid installation.

As illustrated in \cref{fig:water-and-fuel-collection}, there was a notable reduction of $15\pm8$ hours spent collecting water and $36\pm10$ hours spent collecting cooking fuel per 100 households. An increase of 10 kWp in mini-grid PV capacity would lead to $25\pm22$ additional households reporting a decrease in water collection time, and an extra \$50,000 in CAPEX investment would result in an average of $15\pm15$ more households noting a similar improvement.

The proportion of households where men were involved in household chores, like collecting cooking fuel, rose from 10\% to 14\% between the pre-connection and post-connection surveys. However, this increase was less significant in the context of paired households, where the proportion only grew by 0.4 percentage points.

\begin{figure}[t]
	\centering
	\includegraphics[width=0.5\textwidth]{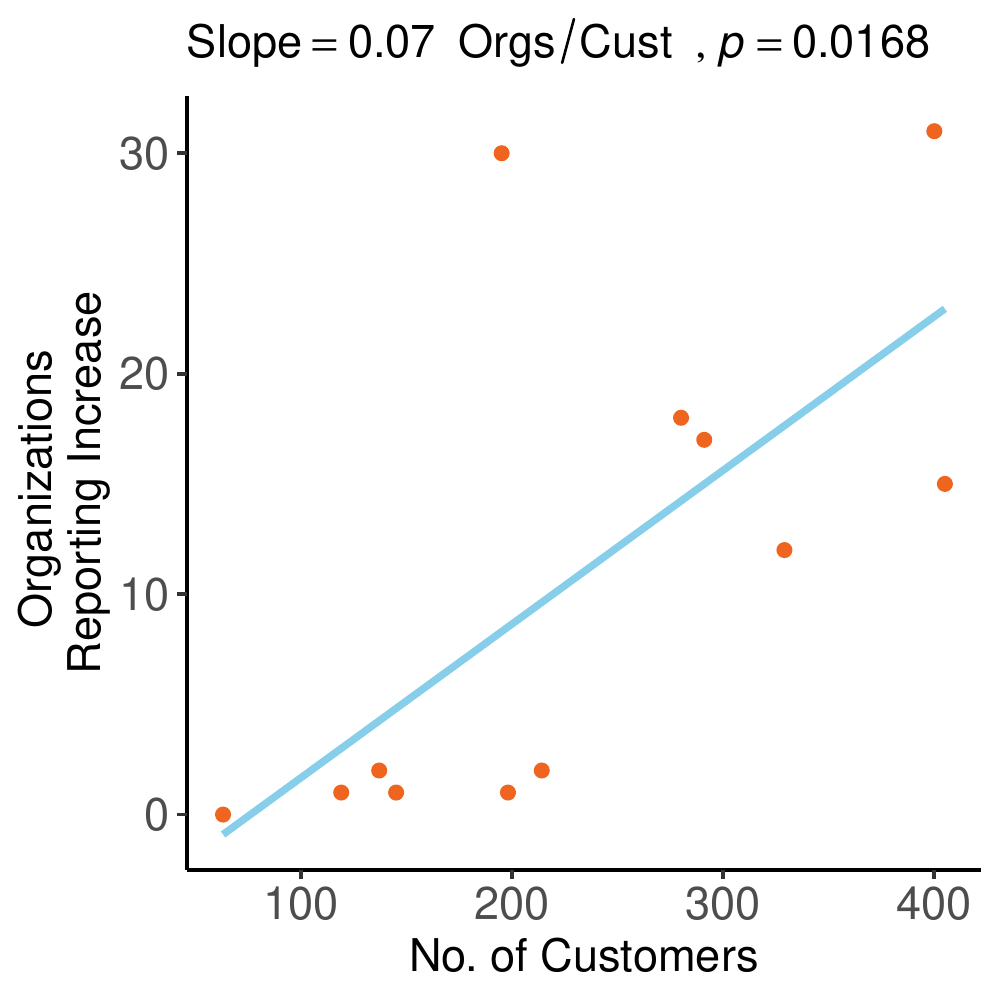}
	\caption{For every ten additional customers connected to a mini-grid, approximately one more business would report hiring at least one more female employee.}
	\label{fig:women_employment}
\end{figure}

The analysis of economic opportunities for women involved two key questions directed at households and organizations focusing on business creation and job availability. Initially, 28\% of households reported having a woman-owned business, but this figure declined to 19\% post-connection. Among paired respondents, this represented a decrease of 4 percentage points. On the other hand, 17\% of the organizations surveyed reported employing at least one new female worker after the introduction of the mini-grid, marking a significant change in this aspect. As illustrated in \cref{fig:women_employment}, it was observed that connecting an additional 100 customers to the mini-grid would lead to $7\pm5$ more organizations employing at least one female worker.

\subsection{Productivity}
\begin{figure}[th]
	\centering
	\includegraphics[width=0.7\textwidth]{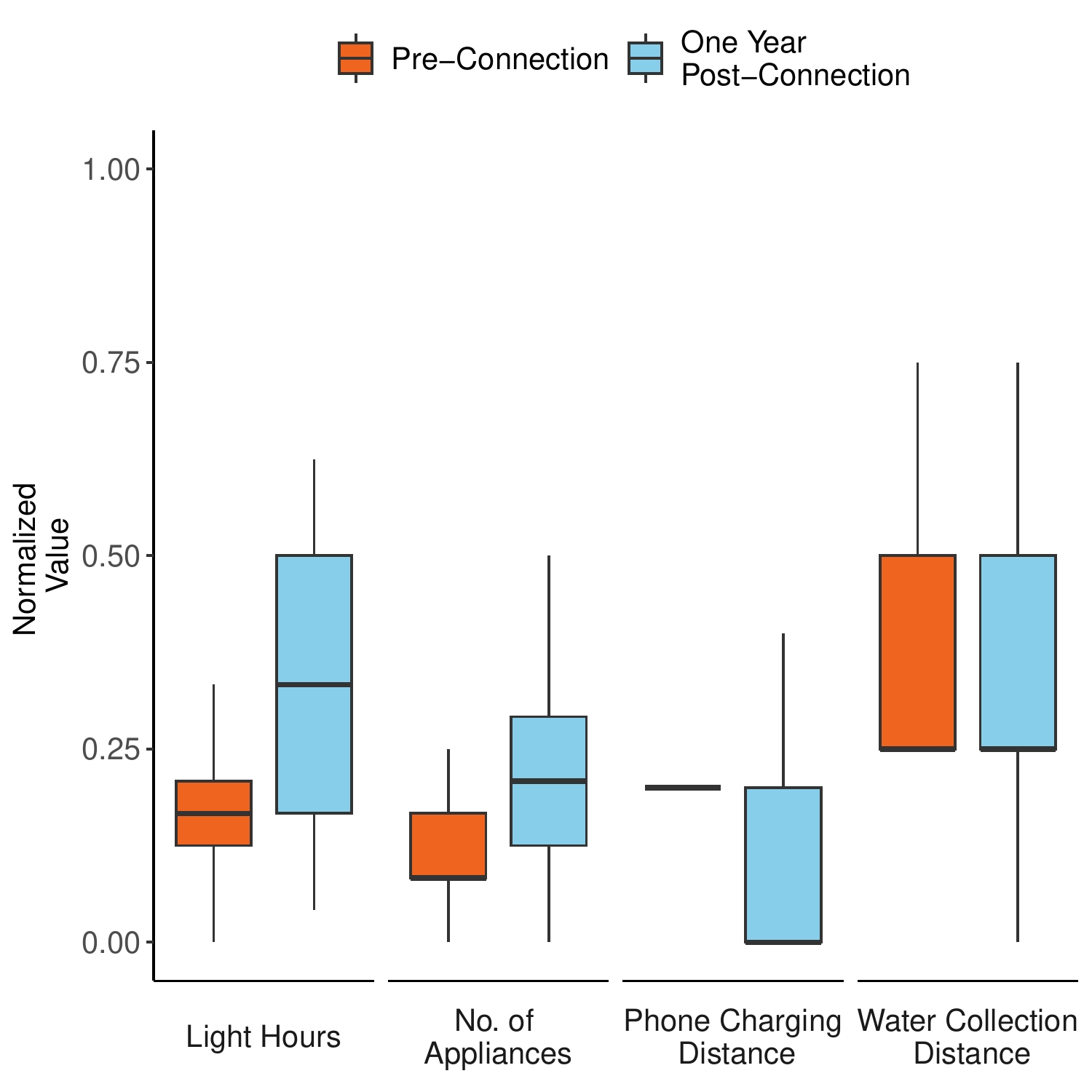}
	\caption{The installation of a new mini-grid had profound effects on community members' productivity, increasing the number of light hours and quantity of electrical appliances in the community while decreasing the distance required for members to charge their phones and collect water.}
	\label{fig:productivity_paired_results}
\end{figure}
The full results of the statistical analysis for questions pertaining to productivity are available in \cref{tab:app:productivity} in the appendix.

As illustrated in \cref{fig:productivity_paired_results}, the installation of the mini-grid also had a notable impact on daily activities, with 58\% of respondents able to charge their phones every day and 94\% doing so at home, reducing the need to travel to neighbors, shops, or other locations for charging by a reported 66\%.

Regarding water and cooking fuel collection time, there was a marked improvement in efficiency post-connection to the mini-grid. In the post-survey, 66\% of respondents reported that it took them less than one hour to collect water, a significant increase from the 57\% who said the same in the initial survey. This increase was also reflected in the paired sample, where an 18 percentage-point change in proportions was observed. Similarly, the time spent collecting cooking fuel decreased notably. In the post-survey, 63\% of respondents indicated that this task took less than one hour, a substantial improvement from the initial survey, where only 35\% reported such efficiency.

\begin{figure}[th]
	\centering
	\begin{subfigure}[t]{0.48\textwidth}
		\centering
		\includegraphics[width=\textwidth]{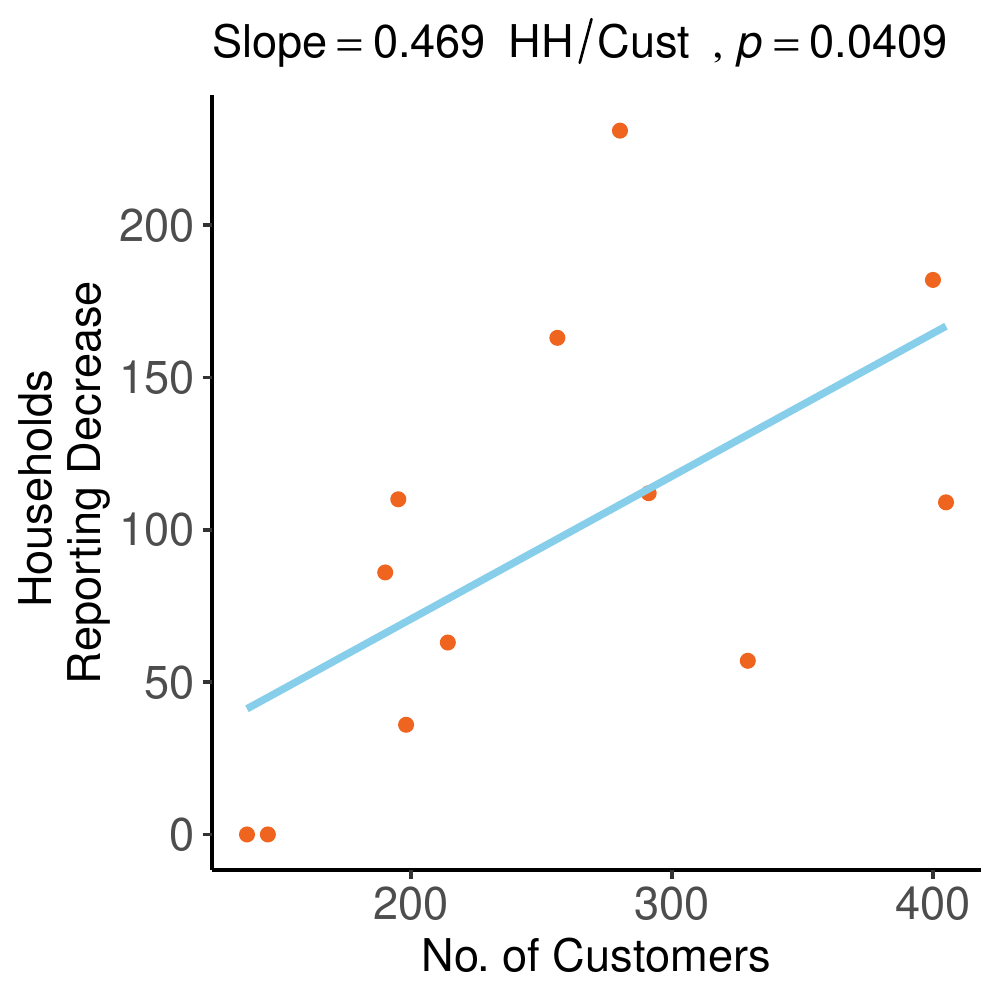}
		\caption{Almost half of all additional community members connected to a mini-grid are statistically likely to reduce or eliminate their usage of unclean power sources such as diesel, petrol, and kerosene.}
		\label{fig:unclean_power_usage}
	\end{subfigure}
	\hfill
	\begin{subfigure}[t]{0.48\textwidth}
		\centering
		\includegraphics[width=\textwidth]{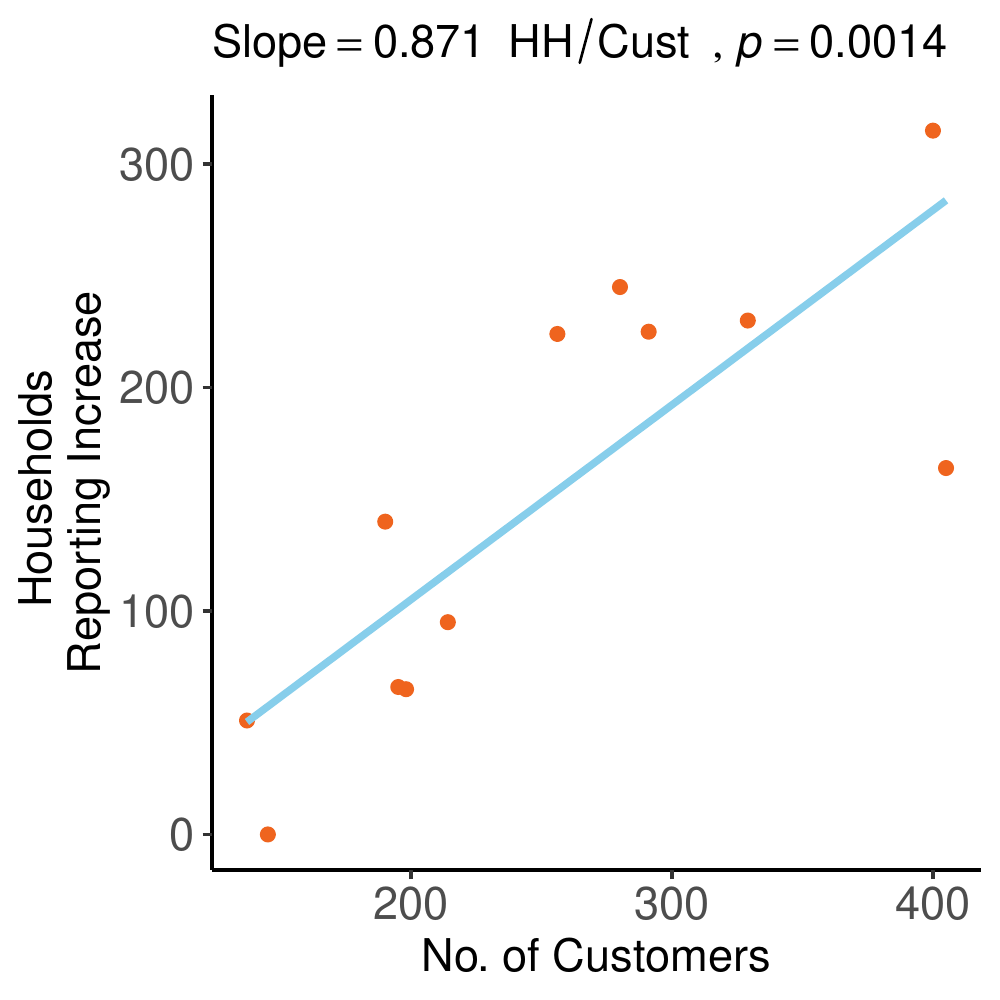}
		\caption{Almost nine out of every ten additional community members connected to a mini-grid are statistically likely to buy new appliances to take full advantage of their new source of electricity.}
		\label{fig:appliances}
	\end{subfigure}
	\caption{Upon connecting to a mini-grid, community members quickly transition from using unclean power sources like diesel, petrol, and kerosene to utilizing this cleaner, cheaper energy option for powering their electrical appliances.}
	\label{fig:appliances-and-unclean-power}
\end{figure}

Prior to connecting to the mini-grid, a majority of the households had limited or no reliable power sources. Specifically, 55\% of households depended on petrol generators. However, one year after the mini-grid connection, this number dramatically decreased to just 1\%, with the mini-grid becoming the primary power source for 91\% of survey respondents. The shift was also evident in the broader reduction of unclean energy sources, including diesel, petrol, and kerosene, which decreased from 66\% in the pre-survey to 19\% in the post-survey. Among paired households, this represented a 7 percentage-point reduction. As illustrated in \cref{fig:unclean_power_usage}, every 100 additional customers connected to the mini-grid would result in $47\pm45$ fewer households using unclean power sources.

Post-connection, households reported significant improvements in their energy access: an average of seven hours of lighting per day and the use of an average of four electronic devices, compared to only four hours of lighting and two appliances before connection. Moreover, 55\% of households acquired new appliances since connecting to the mini-grid. As illustrated in \cref{fig:appliances-and-unclean-power}, for every 100 additional customers connected to the mini-grid, $87\pm45$ acquired new appliances.

\begin{figure}[th]
	\centering
	\begin{subfigure}[t]{0.48\textwidth}
		\centering
		\includegraphics[width=\textwidth]{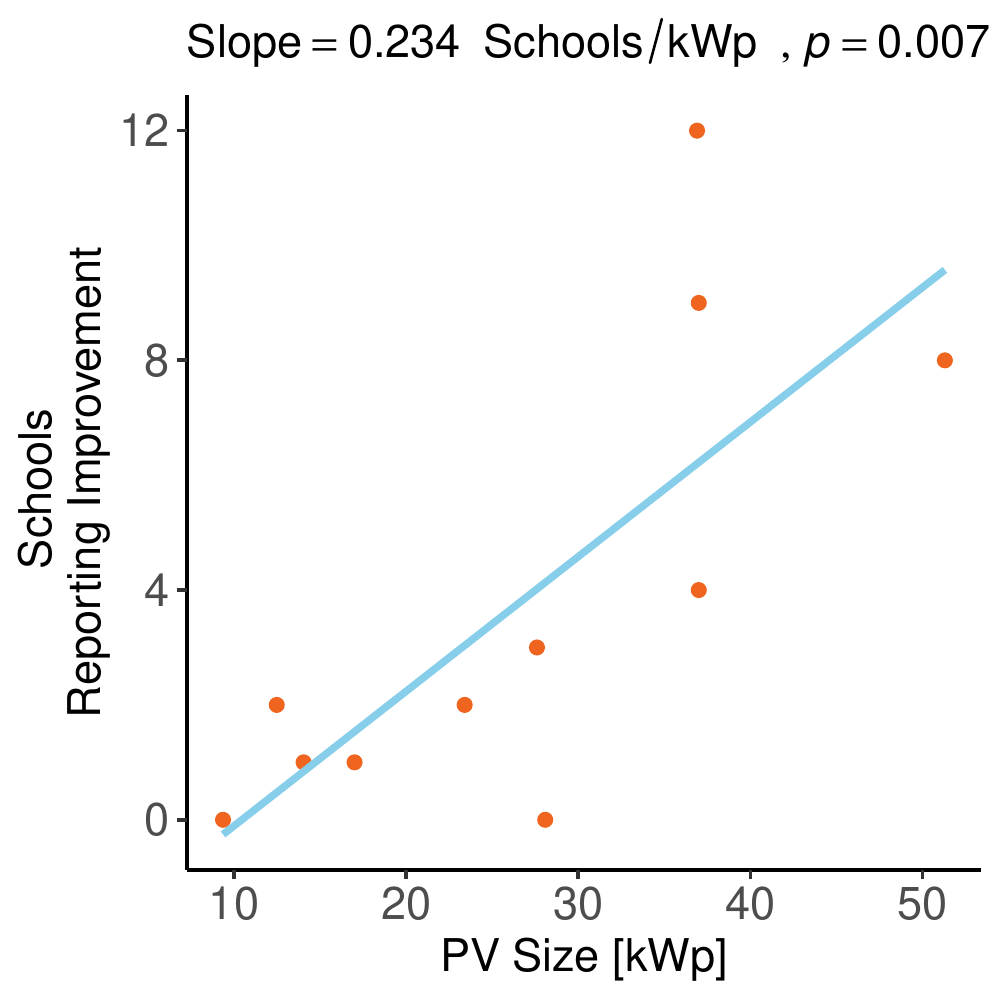}
		\caption{Additional mini-grid solar array capacity resulted in more schools reporting improvements in children's academics.}
		\label{fig:academics_schools}
	\end{subfigure}
	\hfill
	\begin{subfigure}[t]{0.48\textwidth}
		\centering
		\includegraphics[width=\textwidth]{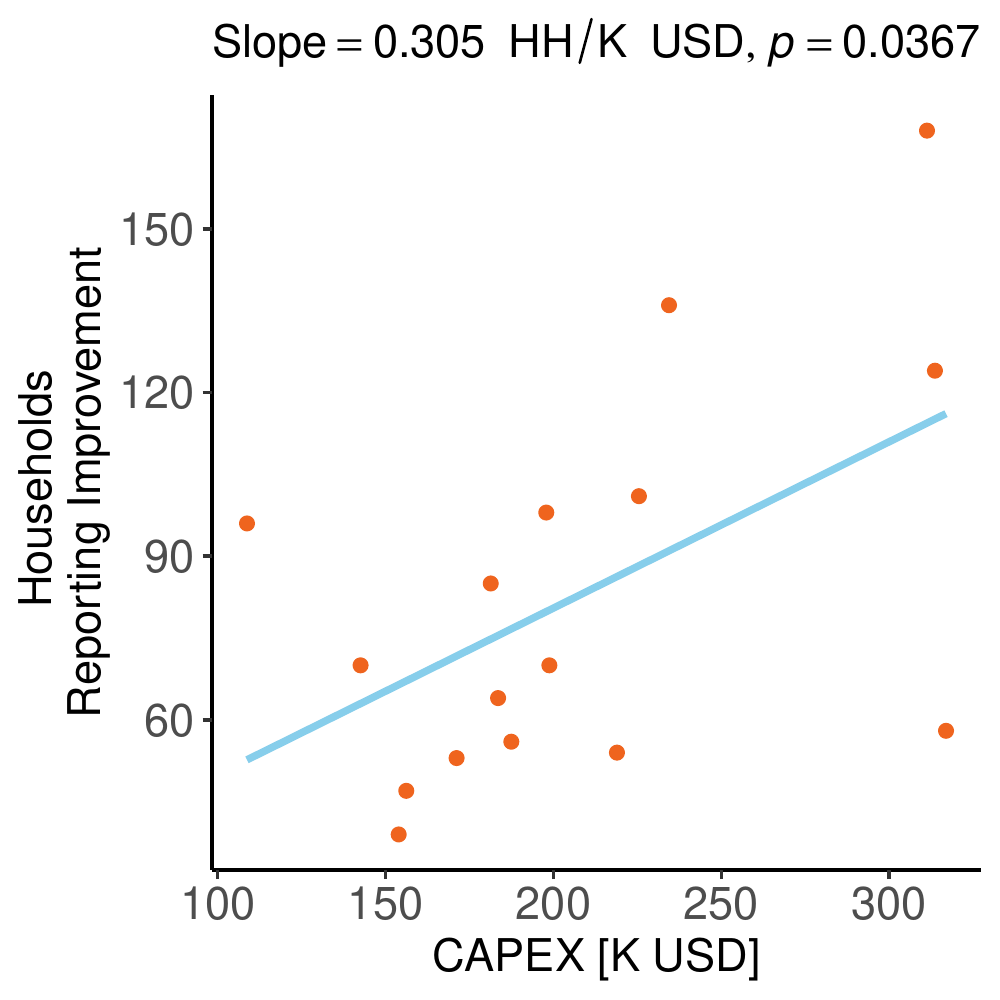}
		\caption{Every additional \$50,000 in CAPEX investment into the mini-grid led to approximately 15 more households reporting an improvement in their kids' academics.}
		\label{fig:academics_households}
	\end{subfigure}
	\caption{Educators and parents both report that larger mini-grids lead to more well-educated students.}
	\label{fig:academic-performance}
\end{figure}

With the connection to the mini-grid providing a more reliable source of power, there were extended lighting hours for both schools and households, directly benefiting school-aged children. In this context, both households and schools were surveyed about changes in academic performance following the mini-grid installation. The results, illustrated in \cref{fig:academic-performance}, were telling: 46\% of households observed a positive change in the academic performance of their children, while a significant 92\% of schools reported similar improvements. Connecting an additional 100 customers to the mini-grid would lead to $19\pm17$ more households observing an improvement in the children's academics. Similarly, an investment of an additional \$50,000 in CAPEX is associated with $15\pm15$ more households reporting a comparable improvement. Furthermore, an increase of 10 kWp in the mini-grid capacity would result in $2\pm2$ more schools noting such an improvement in academic performance.

\begin{figure}[th]
	\centering
	\begin{subfigure}[t]{0.48\textwidth}
		\centering
		\includegraphics[width=\textwidth]{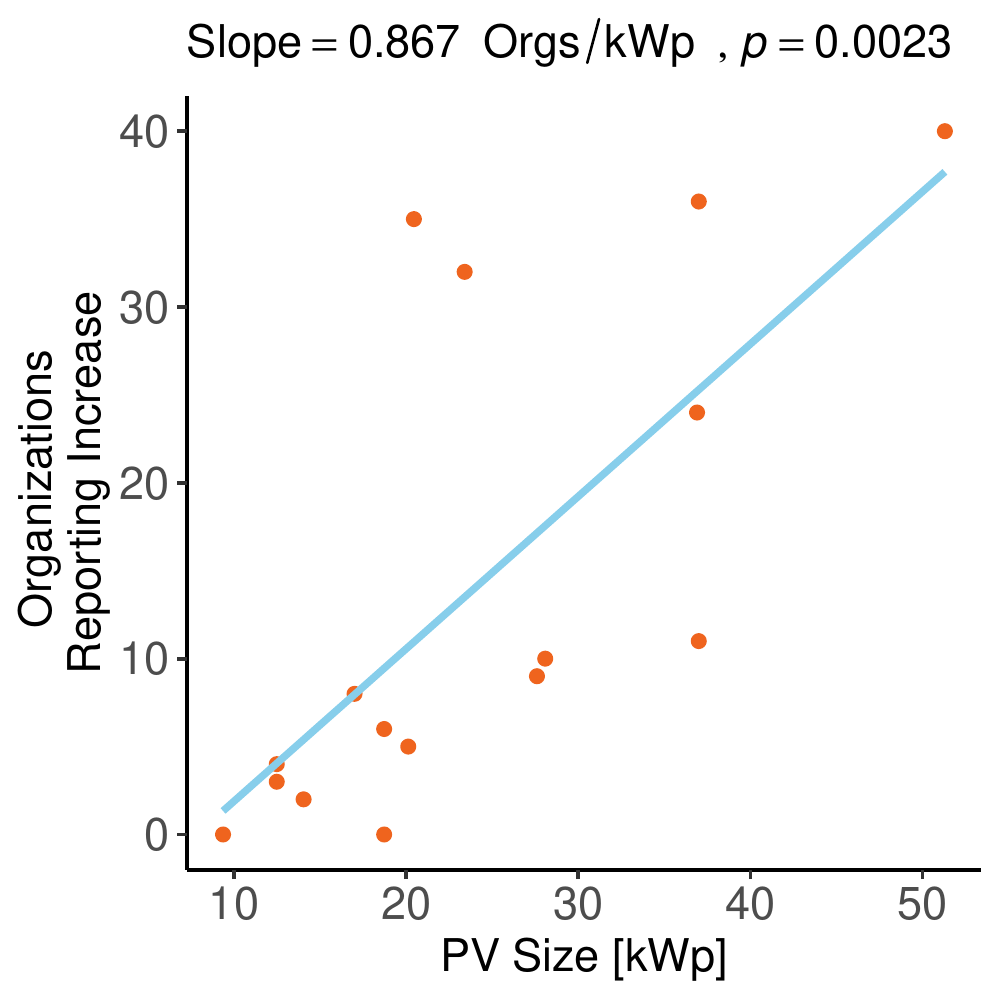}
		\caption{Additional mini-grid solar array capacity resulted in more product or service offerings.}
		\label{fig:ci_offering}
	\end{subfigure}
	\hfill
	\begin{subfigure}[t]{0.48\textwidth}
		\centering
		\includegraphics[width=\textwidth]{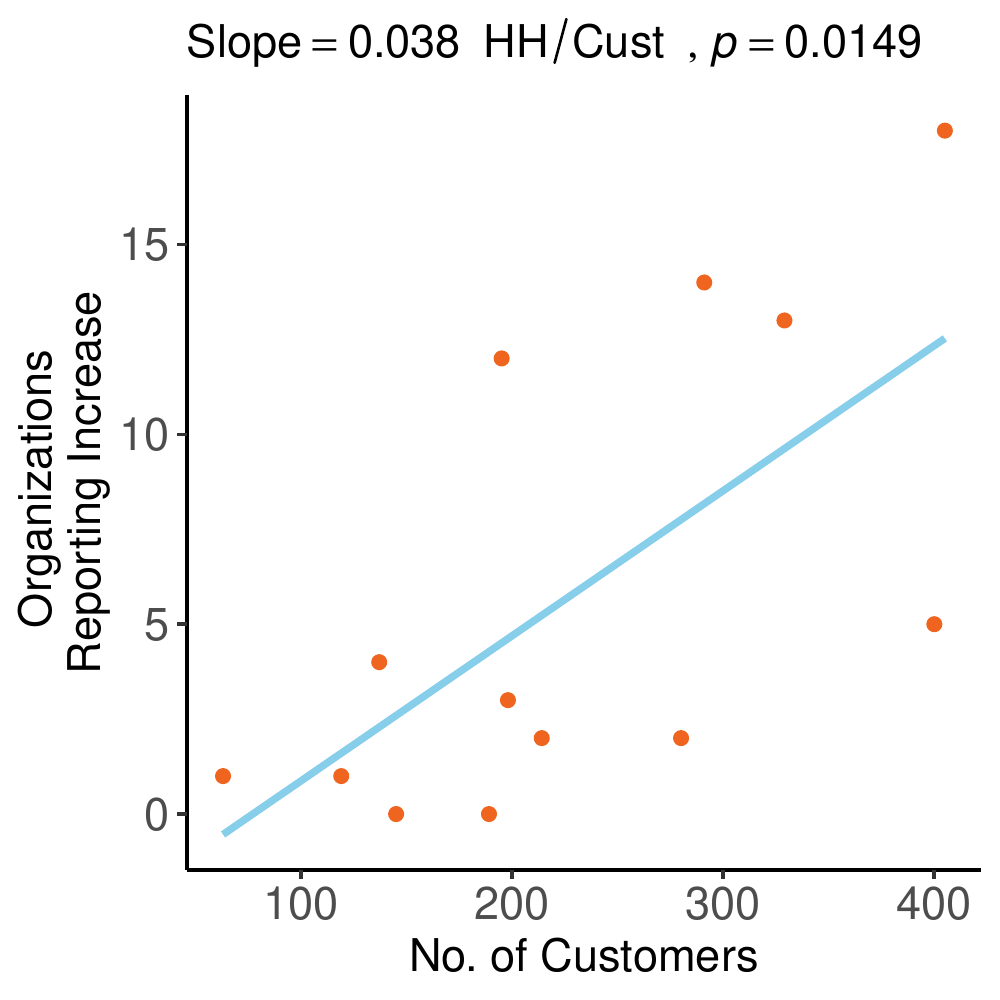}
		\caption{Every additional 100 customer connections led to approximately 4 more establishments hiring at least one more worker.}
		\label{fig:workforce_change}
	\end{subfigure}
	\caption{Establishments reporting an increase in their workforce and offerings.}
	\label{fig:ci_productivity}
\end{figure}

Among commercial and institutional customers connected to the mini-grid, a substantial 87\% reported an increase in their operational hours. Among all establishments, 16\% of them have recruited new employees, while 51\% have broadened their range of products and services. As illustrated by \cref{fig:ci_productivity}, an increase of 10 kWp in the mini-grid capacity is associated with $9\pm5$ more organizations reporting an expansion in their product or service offerings, and every additional 100 customer connections correlates with $4\pm3$ more organizations employing at least one new worker. 

\subsection{Health}
The full results of the statistical analysis for questions pertaining to health are available in \cref{tab:app:health} in the appendix.

Local health clinics in newly electrified communities benefited greatly from access to power from their mini-grids. The access to electricity in clinics showed a significant improvement in the post-survey, with 82\% of respondents indicating that the nearest clinic had electricity, a considerable increase from only 35\% in the initial survey. This positive shift was also evident in the paired sample, where an increase of 53 percentage points was observed. One way in which clinics utilized electricity was to provide refrigeration, enabling the safe storage of vaccines, medications, and blood supplies: 86\% of post-survey responses confirmed refrigeration availability, up from 41\% initially, translating to a 48 percentage-point improvement in the paired sample. Among all the clinics surveyed, 79\% reported the addition or enhancement of at least one service they provide, with a median of three services being improved or added.

\begin{figure}[t]
	\centering
	\begin{subfigure}[t]{0.48\textwidth}
		\centering
		\includegraphics[width=\textwidth]{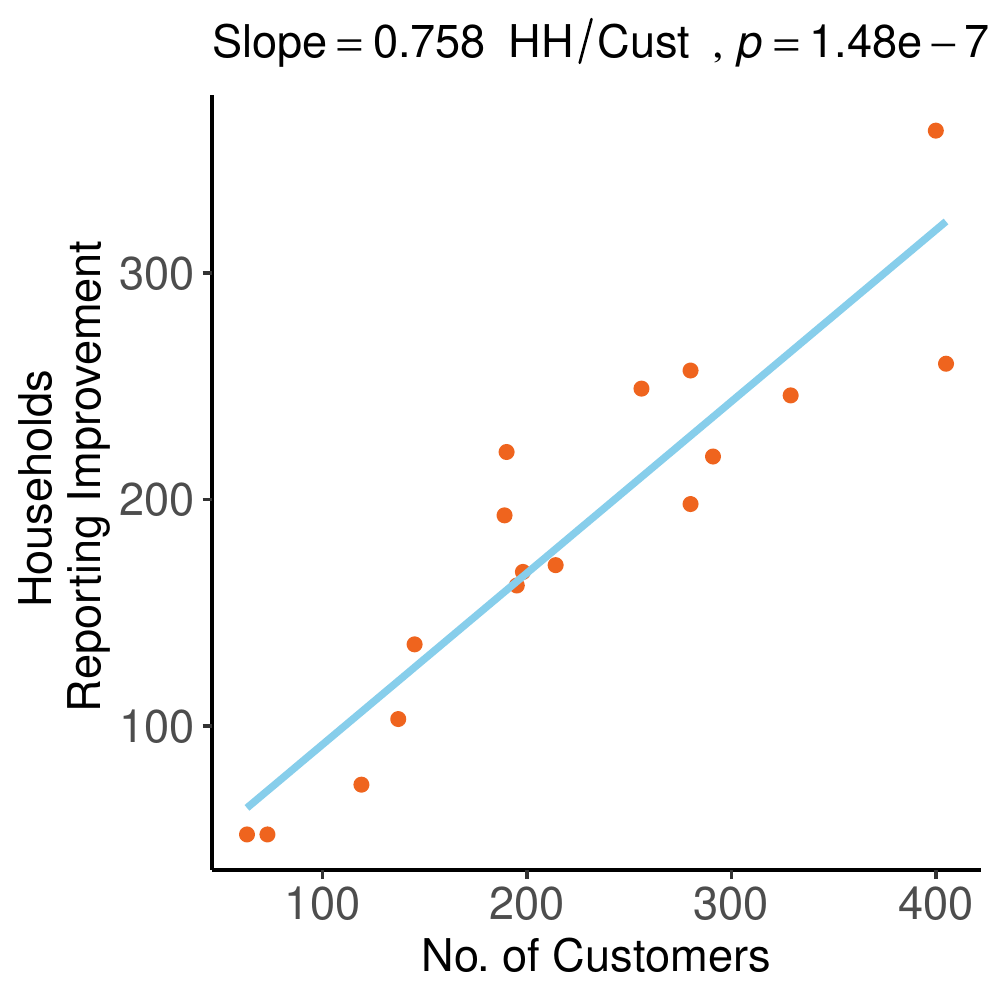}
		\caption{For every four additional customers connected to a mini-grid, approximately three of which will be households, all three will likely report an improvement in the quality of healthcare in their community.}
		\label{fig:better_healthcare}
	\end{subfigure}
	\hfill
	\begin{subfigure}[t]{0.48\textwidth}
		\centering
		\includegraphics[width=\textwidth]{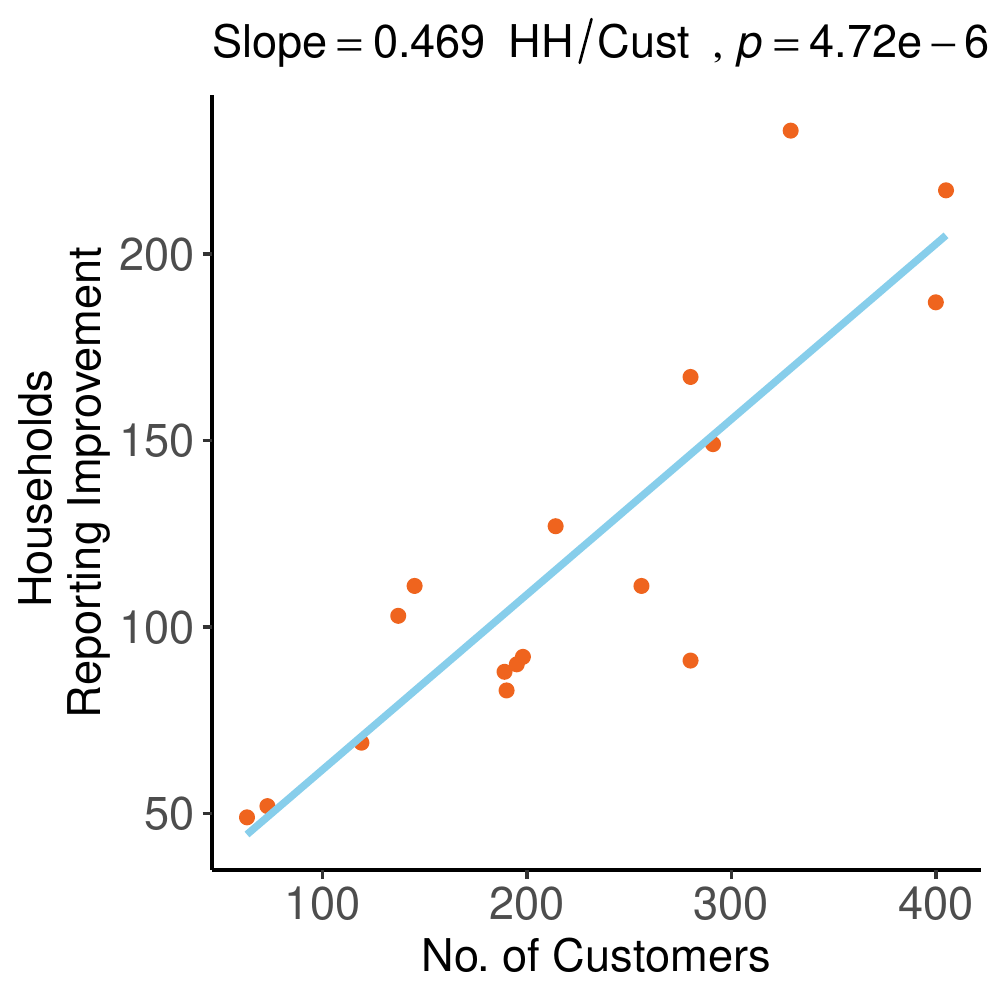}
		\caption{For every four additional customers connected to a mini-grid, two households in the community will likely report an improvement in their overall quality of life.}
		\label{fig:life_improvement}
	\end{subfigure}
	\caption{Connecting additional customers to a mini-grid significantly enhances a community's perceived well-being.}
	\label{fig:well_being}
\end{figure}

Electrification of clinics significantly influenced community members' views on their health. Following the mini-grid installation, 58\% of respondents reported an enhancement in their lives owing to improved healthcare facilities. The connection of an additional 100 customers to a mini-grid results in an average of $76\pm18$ more households reporting access to better healthcare. Furthermore, it leads to approximately $47\pm14$ more households reporting an overall enhancement in their quality of life. These effects are illustrated in \cref{fig:well_being}.

\begin{figure}[th]
	\centering
	\begin{subfigure}[t]{0.48\textwidth}
		\centering
		\includegraphics[width=\textwidth]{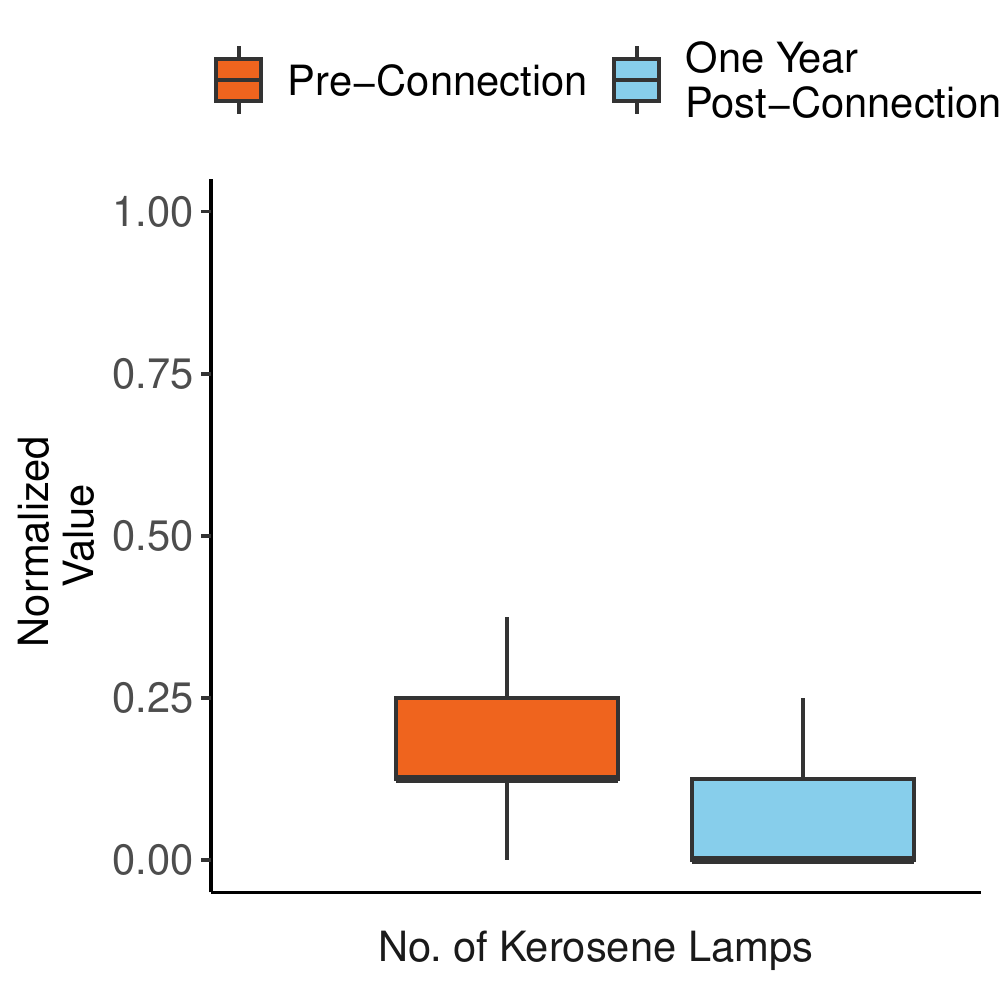}
		\caption{The median household used one kerosene lamp before the mini-grid connection, and none one year after connection.}
		\label{fig:paired-kerosene}
	\end{subfigure}
	\hfill
	\begin{subfigure}[t]{0.48\textwidth}
		\centering
		\includegraphics[width=\textwidth]{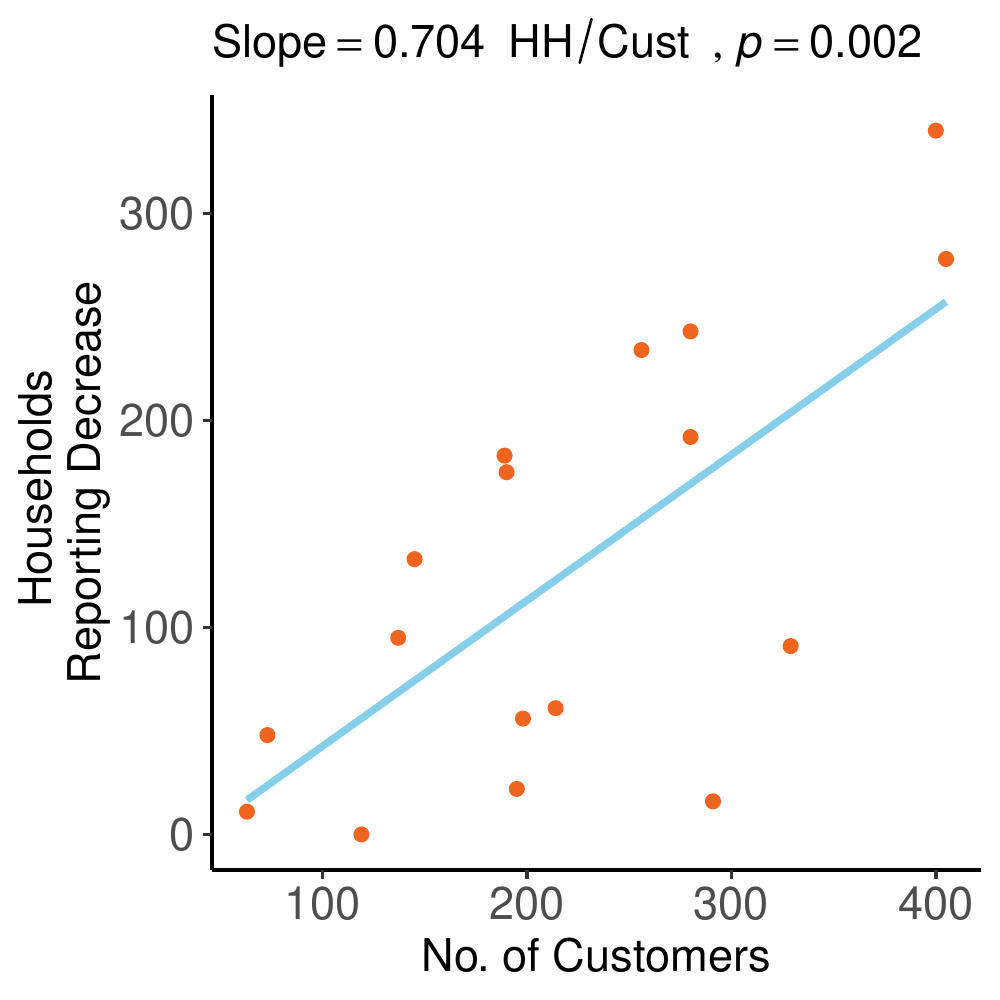}
		\caption{For every ten new customer connections, of which eight will likely be households, seven of those households will likely decrease their reliance on kerosene.}
		\label{fig:regression-kerosene}
	\end{subfigure}
	\caption{Community kerosene consumption decreases significantly after the installation of a new mini-grid.}
	\label{fig:kerosene}
\end{figure}
Connecting 100 additional households to the mini-grid is expected to result in a significant reduction in the use of kerosene lamps: connecting 100 households results in $36\pm20$ fewer kerosene lamps used. \cref{fig:kerosene} illustrates the profound effect of mini-grids on reducing kerosene usage. Put simply, the median household used one kerosene lamp before the mini-grid connection, and none one year after connection.

The installation of mini-grids has led to notable improvements in water quality for the surveyed communities. There was a decrease in the use of dirty water, with pre-survey respondents reporting a 38\% usage rate, which dropped to just 15\% in the post-survey. Concurrently, the majority of households shifted to using community wells or pumps, accounting for 50\% of the water sources post-connection.

When specifically asked about access to clean drinking water, 61\% of post-survey respondents confirmed having such access, a significant rise from 36\% in the initial survey. In the paired samples, this represented a positive increase of 12 percentage points. The primary source of this clean drinking water shifted dramatically, with 60\% of post-survey respondents sourcing it from the community, up from 22\% initially. This shift also led to a reduced reliance on boiling water for purification, which decreased from 41\% to 23\%. One year after connection to a mini-grid, households were 1.9 times more likely to have access to clean drinking water.

\subsection{Safety}
The full results of the statistical analysis for questions pertaining to safety are available in \cref{tab:app:safety} in the appendix.

Although there was no statistically significant change in the overall number of respondents who reported feeling safe in their communities, statistically significant shifts were observed in the reasons cited by those who felt unsafe. The perceived threat from potential theft saw a significant increase, rising from 31\% pre-connection to 73\% one year after connection. Conversely, concerns related to the lack of community lighting decreased substantially from 40\% to 17\%, and the lack of safety due to unsafe travel dropped from 23\% to 7\%. This change is partly attributed to the improvement in community lighting, with 56\% of respondents acknowledging its presence post-connection, a considerable increase from just 22\% pre-connection. This represents a 13 percentage-point increase in the paired sample proportions.

Household lighting saw improvement, with exterior lights increasing from 21\% to 58\% of households post-connection, and 68\% being powered by the mini-grid.

\subsection{Economic Activity}
The full results of the statistical analysis for questions pertaining to economic activity are available in \cref{tab:app:econ} in the appendix.

Following the mini-grid installation, there was a noticeable shift in the occupational dynamics within households. Specifically, 20\% of households reported a change in the occupation of the primary income provider during the first year after connection to the mini-grid. Part of this is due to a discernible increase in entrepreneurial activity. Post-mini-grid installation, 73\% of households confirmed having a business owner in the household, a rise from 70\% observed initially. Notably, 19\% of these households reported that the business was established immediately following the mini-grid installation. This represents a 10 percentage-point increase in business ownership within the paired sample.

\begin{figure}[t]
	\centering
	\begin{subfigure}[t]{0.48\textwidth}
		\centering
		\includegraphics[width=\textwidth]{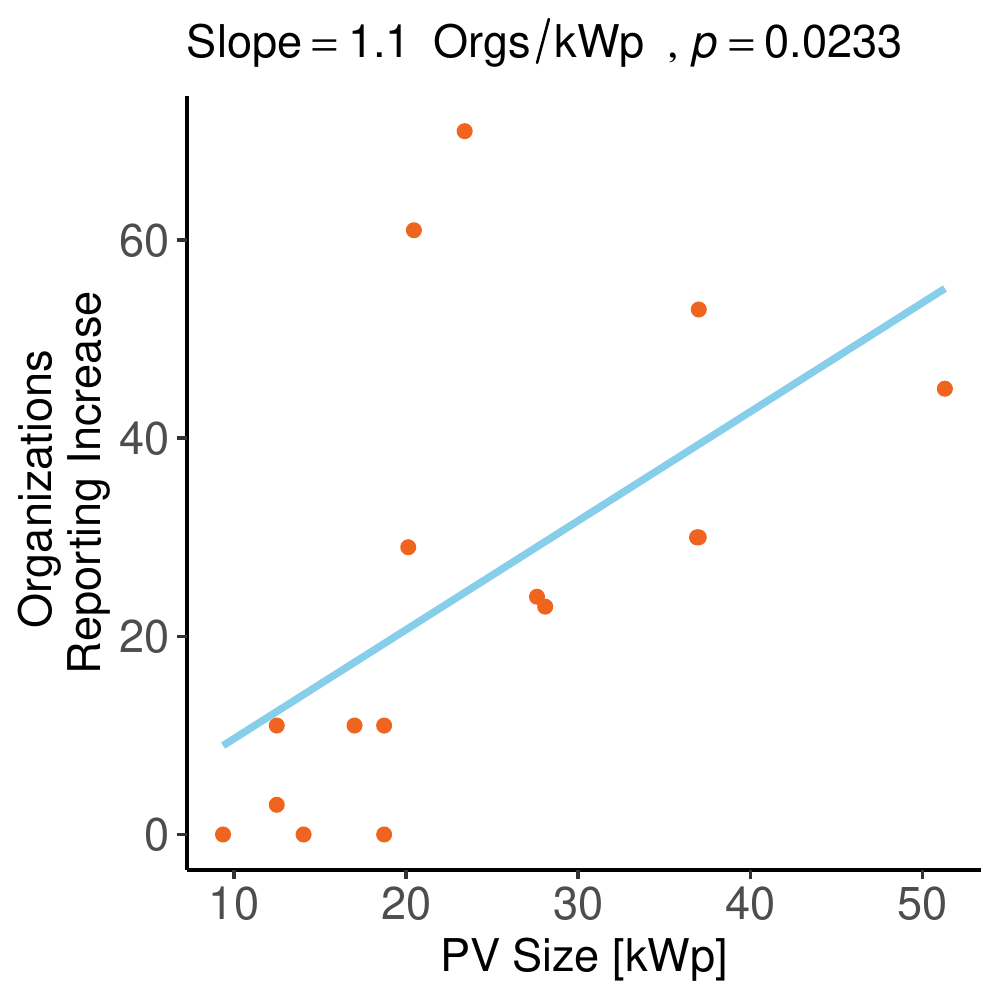}
		\caption{Every additional 10 kWp of installed solar panels correlated with six local businesses reporting increased earnings.}
		\label{fig:earnings_change}
	\end{subfigure}
	\hfill
	\begin{subfigure}[t]{0.48\textwidth}
		\centering
		\includegraphics[width=\textwidth]{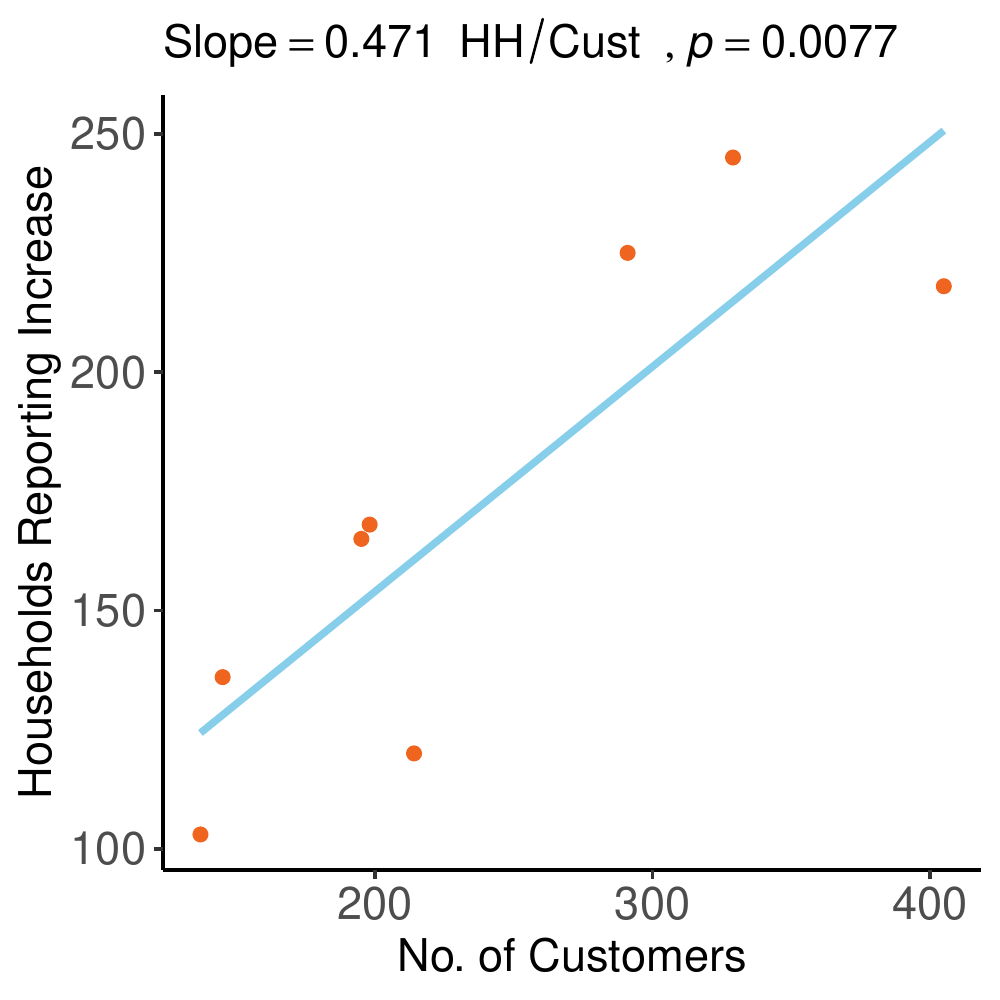}
		\caption{For every ten new customer connections in Kenya, of which eight will likely be households, five of those households will likely experience an increase in their average monthly income.}
		\label{fig:kenya_incomes}
	\end{subfigure}
	\caption{Earnings and Kenyan monthly incomes increased significantly after the installation of a new mini-grid.}
	\label{fig:economic_improvement}
\end{figure}
Among the commercial and institutional customers surveyed, a substantial 85\% reported an increase in their overall earnings following the mini-grid installation. As illustrated in \cref{fig:economic_improvement}, for every additional 10 kWp in mini-grid capacity, there is an increase of $6\pm4$ organizations reporting a boost in their earnings. 

In Kenya, the median monthly income in the communities rose from 5,000 to 6,000 Kenyan Shillings (KES) post-connection, while in Nigeria, the median household income actually decreased from 47,500 to 25,000 Nigerian Naira (NGN) after the mini-grid connection. Notably, 24\% of Kenyan households reported a positive increase in their incomes, compared to 18\% of Nigerian households. Among the paired respondents, the income change was significant: in Nigeria, there was a decrease of $24,079\pm4,496$ NGN, whereas in Kenya, there was an increase of $13,037\pm1,662$ KES. Every 10 additional customer connections in Kenya is correlated with $5\pm3$ more households reporting an increase in their average household income.

\begin{figure}[t]
	\centering
	\includegraphics[width=0.9\textwidth]{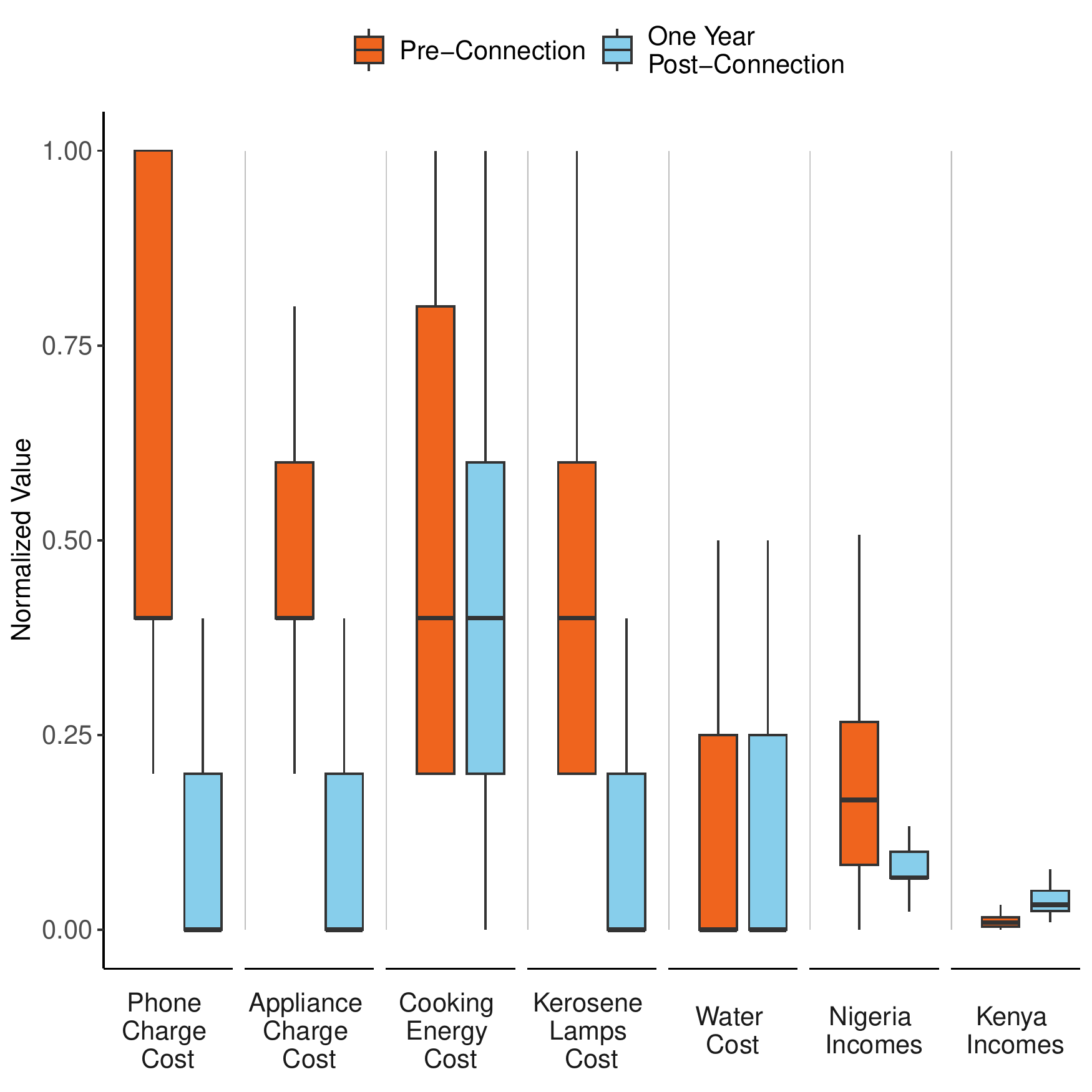}
	\caption{Installing a mini-grid in a community reduced the cost of phone and appliance charging, cooking energy, kerosene lamps, and water.}
	\label{fig:economic_paired_results}
\end{figure}

As illustrated in \cref{fig:economic_paired_results}, before the mini-grid installation, 18\% of Nigerian households spent between 0 and 5,000 NGN monthly on water. Post-installation, 90\% of respondents indicated zero water costs, instead getting all their water from community wells, marking a 94 percentage-point increase in those not paying anything and a substantial decrease of 3,000 NGN per household for water costs. For every 10 customer connections in Nigeria, $2\pm2$ more households would report a decrease in water cost. In Kenya, prior to connection, 74\% of households spent between 0 and 3,000 KES monthly on water, and 27\% paid 5,000 KES or more. After connection, 57\% of households reported no water expenses.

Before connecting to the mini-grid, 84\% of Nigerian households were spending between 0 and 1,000 NGN monthly to charge their phones due to the absence of electricity. Post-connection, a significant reduction in this expense occurred, with 96\% of post-survey respondents reporting no additional costs beyond their household power bill. In the paired sample, this equates to a 97 percentage-point increase in those not paying for phone charging and a marked decrease of $900\pm100$ NGN per household each month. For every additional 10 customer connections in Nigeria, $7\pm4$ more households would report a decrease in phone charging cost. In Kenya, pre-connection, 43\% of households spent between 0 and 750 KES monthly on phone charging, and 55.5\% paid 1,000 KES or more. After connection, a considerable number of households saw this cost alleviated, with 45\% of post-connection respondents indicating no extra expenses beyond their household power bill. In the paired sample, this translates to a 15 percentage-point increase in those not incurring phone charging costs and a significant monthly reduction of 100 KES per household.

Initially, 84\% of Nigerian households incurred costs between 0 and 6,000 NGN monthly for appliance charging due to unreliable power sources. However, one year after connecting to the mini-grid, 93\% of respondents no longer faced any additional charging costs. In the paired sample, there was a 94 percentage-point increase in households reporting no appliance charging costs and a significant monthly reduction of 3,000 NGN per household. In Kenya, 99\% of households initially spent between 0 and 4,000 KES monthly to charge appliances. A year after mini-grid connection, 37\% of households reported no longer having this expense. The paired sample showed a 13 percentage-point increase in those not paying for appliance charging and a notable decrease of 150 KES per household each month.

In Nigeria, before the mini-grid installation, 84\% of respondents spent between 0 and 3,000 NGN monthly on cooking energy. One year after installation, 34\% reported their cooking costs had fallen to between 0 and 1,000 NGN, a significant increase from 14\% pre-connection. In the paired sample, this shift represented a 26 percentage-point increase in those with a lower cooking energy bill in the 0 to 1,000 NGN range, alongside a notable monthly decrease in cooking energy costs of 1,500 NGN per household. In Kenya, prior to the mini-grid connection, 96\% of households were spending between 0 and 2,000 KES monthly on cooking energy. Post-connection, a cost reduction was observed, with 79\% now paying either nothing or between 0 and 1,500 KES. The paired sample analysis showed a 4 percentage-point increase in households paying nothing, a 27 percentage-point decrease in those spending between 0 and 1,000 KES, and a 17 percentage-point increase in those now spending between 1,000 and 1,500 KES on cooking energy.

Prior to the mini-grid connection, 57\% of Nigerian households were spending between 0 and 1,400 NGN monthly on expenses related to kerosene lamps. A year after connection, a significant change was observed: 91\% of post-survey respondents reported no expenditure on kerosene. In the paired sample, this equated to an 89 percentage-point increase in households no longer incurring costs for kerosene lamps, with a substantial monthly reduction of $1,000\pm400$ NGN per household. For every additional 10 customer connections in Nigeria, $6\pm2$ more households would report a decrease in the amount spent on kerosene lamps. In Kenya, initially, 98\% of households spent between 0 and 1,000 KES on kerosene lamp-related costs. After a year of mini-grid connectivity, 72\% of respondents reported eliminating these expenses. This change represented a 47 percentage-point increase in the paired sample for those not needing to spend on kerosene lamps, and a notable monthly decrease of 200 KES per household.
\section{Discussion and Qualitative Insights}
\label{sec:discussion}
This study conducts a thorough analysis of solar mini-grids' impact in surveyed communities, examining changes across a wide range of variables in the first year post-installation. Utilizing both descriptive and inferential statistics, it highlights key observed changes. Although not exhaustive, the study captures diverse socio-economic benefits of electrification, shedding light on its potential positive impact on African rural communities. The findings are organized into themes: gender equality, productivity, health, safety, and economic activity, enabling detailed exploration of implications and study limitations. This approach aims to comprehensively understand the multifaceted impact of solar mini-grid implementation.

In this context, gender equality was defined as increased equal access to education for both boys and girls, enhanced economic opportunities for women, and reduced time spent on household chores typically assigned to young and adult females in those communities. There were 11 questions pertaining to this KPI from which several variables were extracted for analysis. The results indicate a notable and significant improvement ranging from access to education for boys and girls alike to increased economic inclusion for women. The associated investment into the installation and capacity of the mini-grid proved to be positively correlated with higher school enrollment rates across communities.
Responses from open-ended survey questions added depth to these findings. Post-connectivity, school-aged children, especially girls, reported higher grades, attributed to extended study hours into the night and the ability to complete schoolwork more effectively. Additionally, the mini-grid connection facilitated access to the internet, providing children with additional learning materials. This connectivity proved particularly beneficial for girls, who gained increased exposure to and knowledge about feminine care and hygiene issues.
Furthermore, the reduction in time spent on chores like water and cooking fuel collection offered relief to the primary caregivers in the household, who were often young or adult women. Furthermore, the availability of nighttime lighting allowed for the completion of chores later in the evening, resulting in earlier meal preparations. The broader economic landscape also saw a transformation, with the mini-grid connection opening up new job opportunities from which women substantially benefited.

Before the mini-grid connection, many economically productive activities were often determined by the natural light hours and were limited by little-to-no access to electricity. For instance, many residents had to travel to charge electronic appliances such as phones and, so, restrict the time they could spend on other activities.
Respondents involved in the fishing industry noted that their fishing activities were limited due to the absence of ice for preserving their catch before reaching shore. Consequently, they restricted their daily catch to quantities they could sell immediately, as smoking was the only available method for preservation.
These inefficiencies were curtailed or eliminated by a reliable power source. More broadly, productivity is related to light hours for households, time spent on and distance traveled for specific activities like water collection and appliance charging, and enhanced operational hours and workforce capacity for organizations. The surveys contained 31 productivity-related questions for households and 7 for businesses. Analyses of the metrics related to productivity demonstrated gains and savings in terms of resources allocated to various activities, thus pointing to greater efficiency at the individual and community levels. Connecting to the mini-grid gave more light hours, reduced reliance on unclean or non-renewable power sources, and provided on-demand charging stations at home for all their electronic devices. This result affirms the conclusions of \cite{wassie2021socio, uwineza2021analysis}, where the installations of solar panels and a small hydroelectric power plant in Ethiopian and Rwandan communities, respectively, led to increased employment, educational attainment, and business activity. Our study expands upon this result by examining a broader set of questions that capture economic productivity both at the household and business level.
The responses from open-ended survey questions provided additional depth and nuance to these findings. The availability of electronic information devices such as televisions and radios has expanded access to information for residents, while the convenience of charging phones has enhanced communication within families and the community at large. The ability to access social media platforms has also heightened residents' awareness of local, national, and international news. Furthermore, the acquisition of electric appliances like clothes irons has enabled residents to wear ironed clothes, contributing to an improved sense of self-presentation.
There was also an improvement in the academic performance of school-aged children affirmed by both households and schools. Children could now extend their study hours at home, and schools could provide more services to support the student body, potentially at later hours. Most organizations also had a notable increase in their hours of operation---now with a reliable source of electricity---thus driving more business growth in hiring more workers and expanding their lines of products and services.

Sixteen health-related questions were posed to households, focusing on the ease of access to the nearest clinic, use of kerosene lamps, water purification methods, and the overall quality of healthcare. The results highlighted a significant positive impact on the health and well-being of communities connected to mini-grids. Residents gained access to clean drinking water, primarily through community wells or pumps, reducing health risks associated with the use of unclean water.
Clinics reported shorter wait times, extended hours of operation, the capacity to treat more patients, and improved cold storage for vaccines and medicines. Residents directly confirmed an overall enhancement in healthcare services and living conditions.
The adoption of mini-grids led many households to discontinue the use of hazardous kerosene lamps, known for their risks of poisoning, fires, and explosions. Here, too, open-ended survey responses further enriched the findings. The consistent voltage supply from the mini-grid has enabled a shift away from petrol-powered generators, significantly reducing noise and pollution in residential areas. Residents, like clinics, also benefit from cold storage facilities, allowing access to cold drinking water and other chilled beverages, alongside a notable reduction in food wastage. This transition has contributed to an improved quality of life for the residents, underscoring the multifaceted health and lifestyle benefits brought about by the mini-grid installation.

The safety of residents was assessed based on how safe they felt and any potential reasons they might feel unsafe, including due to theft, unsafe travel, and lack of community lighting, with 10 questions allocated from the surveys. Households were more likely to have exterior lighting, thus contributing to a feeling of safety in and around the house. On the feeling of safety, there was no significant increase or decrease, thus indicating that most residents did not affirm feeling more or less safe. However, there was a notable and positive change in the reasons related to electrification. Indeed, there was a drop in respondents denoting community lighting and lack of safety when traveling as reasons for feeling unsafe. Even without an overall increase in the feeling of safety, those two reasons indicate that the mini-grid installation reduced the incidence of electrification-related concerns. On the other hand, there was a significant increase in the perceived threat of theft following the mini-grid installation. This observation suggests a need for further investigation to assert a direct correlation with the mini-grid installation. Future work should avoid this limitation by expanding the number of questions dedicated to assessing safety.

Survey questions measuring economic activity assessed changes in household incomes and spending, earnings for businesses, employment, and business ownership. This section contained 17 questions. The median household income in Nigeria decreased between 2021 and 2023, but this can likely be best explained by macroeconomic trends \cite{wb-nigeria-2,wb-nigeria-1}. On the other hand, Kenyan households observed an increase in income after mini-grid connection.
Among the paired respondents---those community members who were present both before the mini-grid installation and at least until the post-survey one year after connection---there was a significant rise in median monthly income. This group saw their median income increase from 4,000 KES to 18,000 KES, highlighting a substantial economic uplift due to the mini-grid installation. This is in contrast to the findings of \cite{lee2020experimental}, which did not find meaningful impacts on economic and non-economic outcomes after the expansion of electric grid infrastructure in rural Kenya, potentially due to the low energy utilization they observed at the household level.

Across both countries, business ownership was higher, and previously established businesses had increased earnings. The mini-grid installation bolstered the entrepreneurial economy as electrification brought additional economic opportunities. Households could also save on the charging cost for phones and appliances, cooking energy, and water costs, freeing up money for other investments.
\section{Conclusion}
\label{sec:conclusion}
This paper offers an in-depth evaluation of the tangible socioeconomic impacts of solar mini-grids in rural communities across sub-Saharan Africa. It utilizes robust empirical methods to examine five Key Performance Indicators: gender equality, productivity, health, safety, and economic activity. This constitutes the first comprehensive analysis of the social and economic effects of solar mini-grids in rural African settings.

Surveys were conducted in Nigeria and Kenya at selected sites before and one year after solar mini-grid installation, yielding a detailed insight into the mini-grids' impacts. Generating both quantitative and qualitative data, the surveys provided a holistic perspective on various aspects such as household chores, education access, time use, healthcare, and incomes. Descriptive statistics and comparative tests, including the paired \textit{t}-test and McNemar's test, were used to evaluate significant differences pre- and post-installation.

The study showed marked improvements in children's schooling, including better academic performance and less involvement in household chores, indicating that mini-grids enhanced educational opportunities and potentially raised literacy rates. Additional lighting hours and reduced time for water and cooking fuel collection from the mini-grid led to increased productivity. Additionally, replacing hazardous kerosene lamps with the cleaner, reliable power of the mini-grid improved both community health and productivity.

The mini-grid's on-demand home electricity significantly reduced household expenses, notably in services like phone charging. This installation led to diverse improvements, positively affecting many aspects of community life.

Despite the study's positive findings, certain limitations were encountered. The survey's structure led to an imbalance between discrete and continuous variables, limiting the use of regression analyses. This restriction affected the robustness of testing electricity consumption effects at the household level. Site-level analyses provided some compensation, but they offered only community-wide results, amalgamating individual household and business impacts. Furthermore, many survey questions yielded outputs unsuitable for parametric hypothesis testing due to their lack of meaningful order or magnitude.

This study's insights on solar mini-grids in Nigeria and Kenya pave the way for future research. Longitudinal studies tracking participants over years would deepen understanding of mini-grids' long-term effects. Comparing these communities with nearby unelectrified ones could create a natural quasi-experiment. Broadening the study to other regions would enhance understanding of mini-grids' diverse impacts. Assessing how evolving renewable technologies affect these impacts, and examining the influence of government policies, subsidies, and international aid, are also crucial. Addressing these aspects will enrich the knowledge base, aiding the effective use and optimization of solar mini-grids for greater social and economic advantages.

Solar mini-grids have significantly driven positive transformations in communities, leading to enhanced opportunities for women and girls, better healthcare, and economic growth. Their impact on various socioeconomic factors highlights their role in achieving the Sustainable Development Goals by 2030.

\iftoggle{anonymous}{}{
	\section*{Conflict of Interest Statement}
This study was conducted with the involvement of several authors who have direct employment ties with the primary funder, Renewvia Energy Corporation. Specifically, authors A.T. Carabajal, A. Orsot, and N.S. Selby are employed by Renewvia Energy Corporation. Additionally, G.T. Jarrard III holds the position of CEO at Renewvia Energy Corporation.
	\section*{Funding Statement}
This project was primarily funded by Renewvia Energy Corporation. While no specific grants were allocated for this project, authors employed by Renewvia Energy Corporation led the study design, analysis, interpretation, and writing of the report. This collaboration was in partnership with the African Leadership University; the University of Nairobi; and the University of California, Berkeley. Data collection was carried out by independent, third-party surveyors.
}
\section*{Data Availability Statement}
In line with our commitment to transparency and scientific reproducibility, the anonymized responses from the surveys conducted for this study are released alongside the publication of this paper. We have ensured that all personally identifiable information has been removed from these responses to maintain the confidentiality and privacy of the participants. Per the request of the communities, this includes redacting the names and precise locations of the communities involved, replacing them with the state in which each community resides (e.g, ``Cross River 1,'' ``Turkana 2,'' etc.). This measure is crucial to protect the identity of the communities while still providing context and geographical relevance to the data. This anonymization process was undertaken with great care to preserve the integrity and utility of the data for further analysis and research.

Researchers and interested parties can access these datasets to validate our findings, conduct further analysis, or for comparative studies in similar fields. The data is available in CSV format.

By providing open access to our research data, we aim to foster a collaborative and transparent research environment that encourages knowledge sharing and collective advancement in our field.

\section*{References}
\bibliographystyle{iopart-num}
\bibliography{main}
\newpage
\section{Appendix}
\label{sec:appendix}

\begin{figure}[ht]
	\centering
	\includegraphics[width=\textwidth]{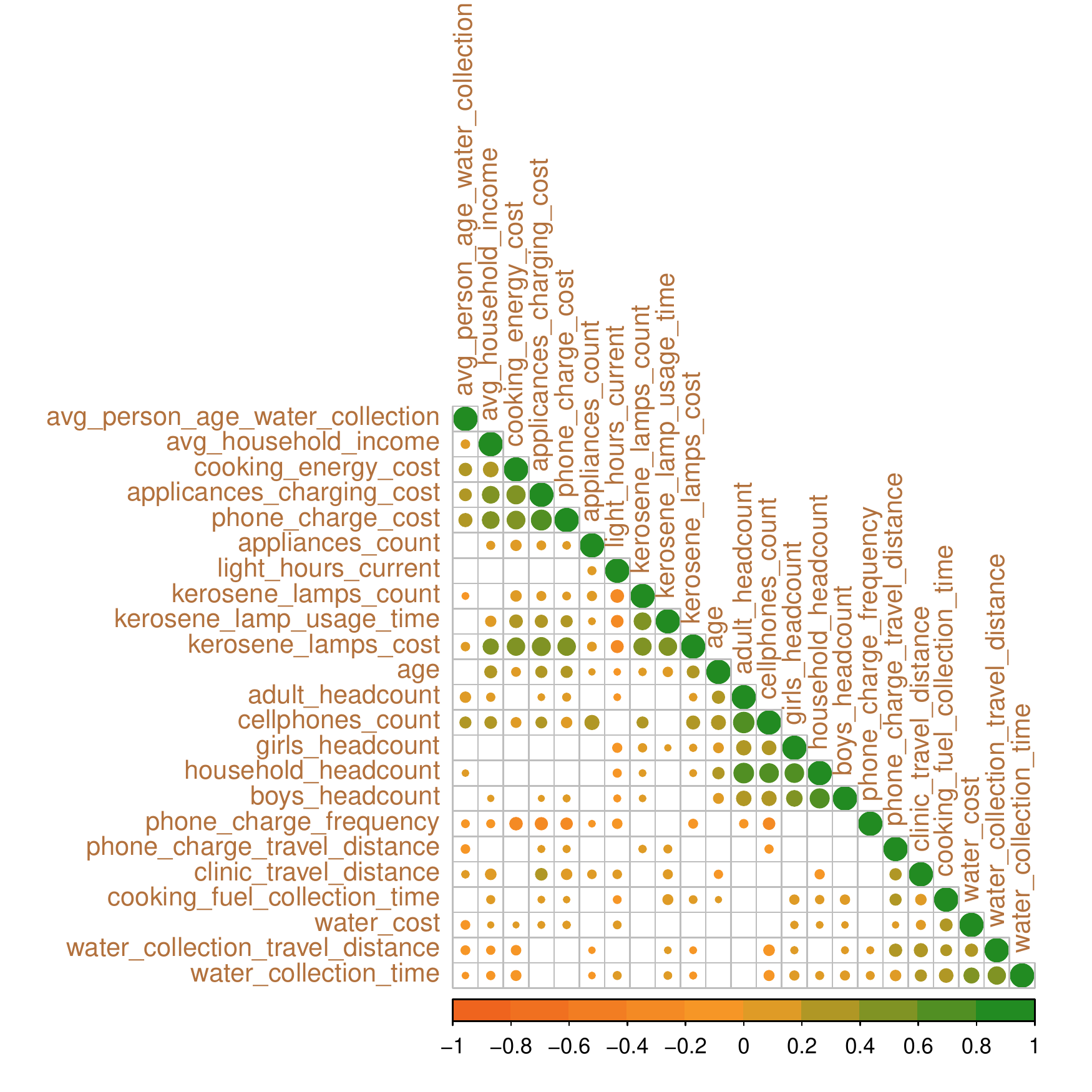}
	\caption{Correlation Matrix of Household Pre-Survey Responses to 23 Numerical Questions}
	\label{fig:corr-pre}
\end{figure}

\begin{figure}
	\centering
	\includegraphics[width=\textwidth]{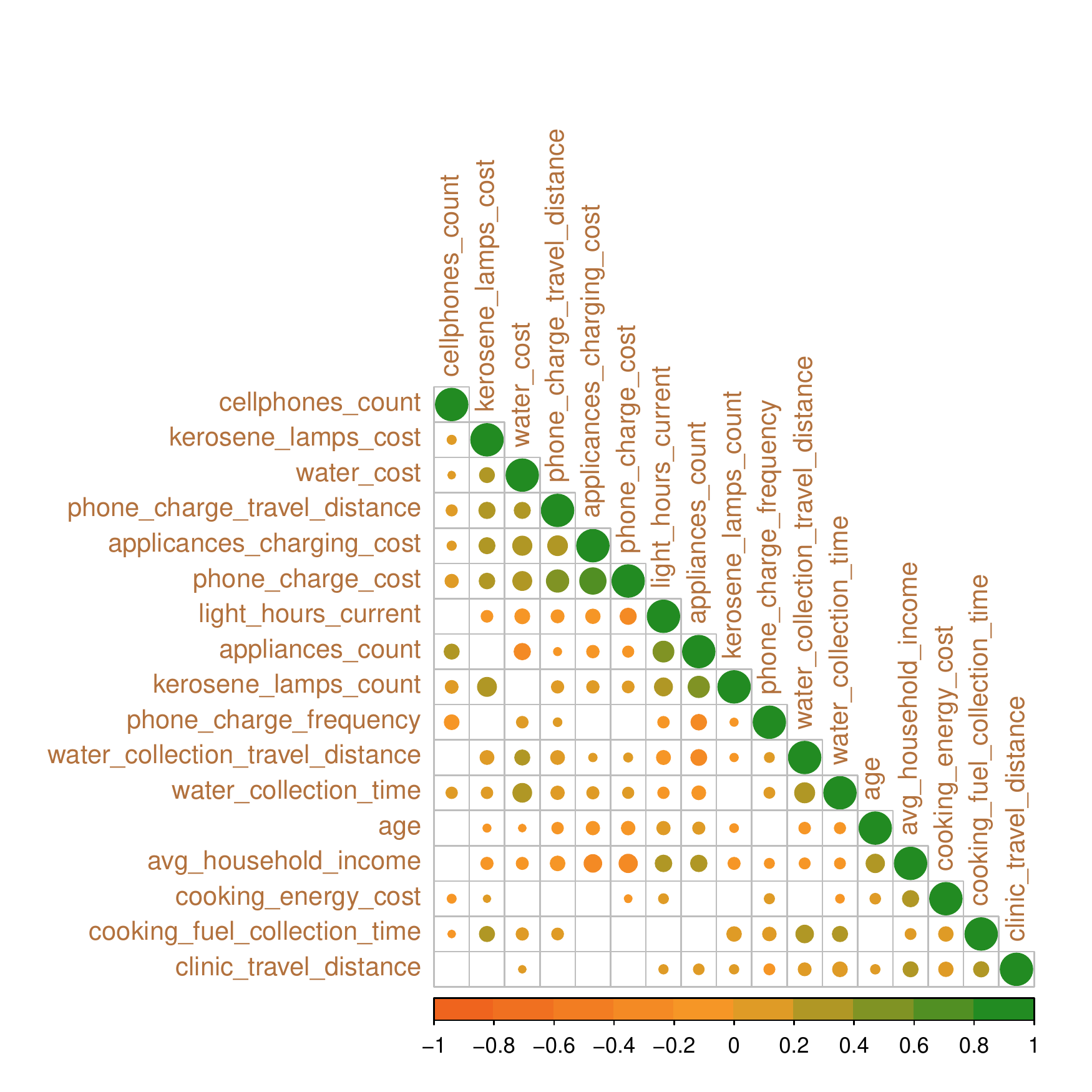}
	\caption{Correlation Matrix of Household Post-Survey Responses to 17 Numerical Questions}
	\label{fig:corr-post-hh}
\end{figure}

\begin{figure}
	\centering
	\includegraphics[width=\textwidth]{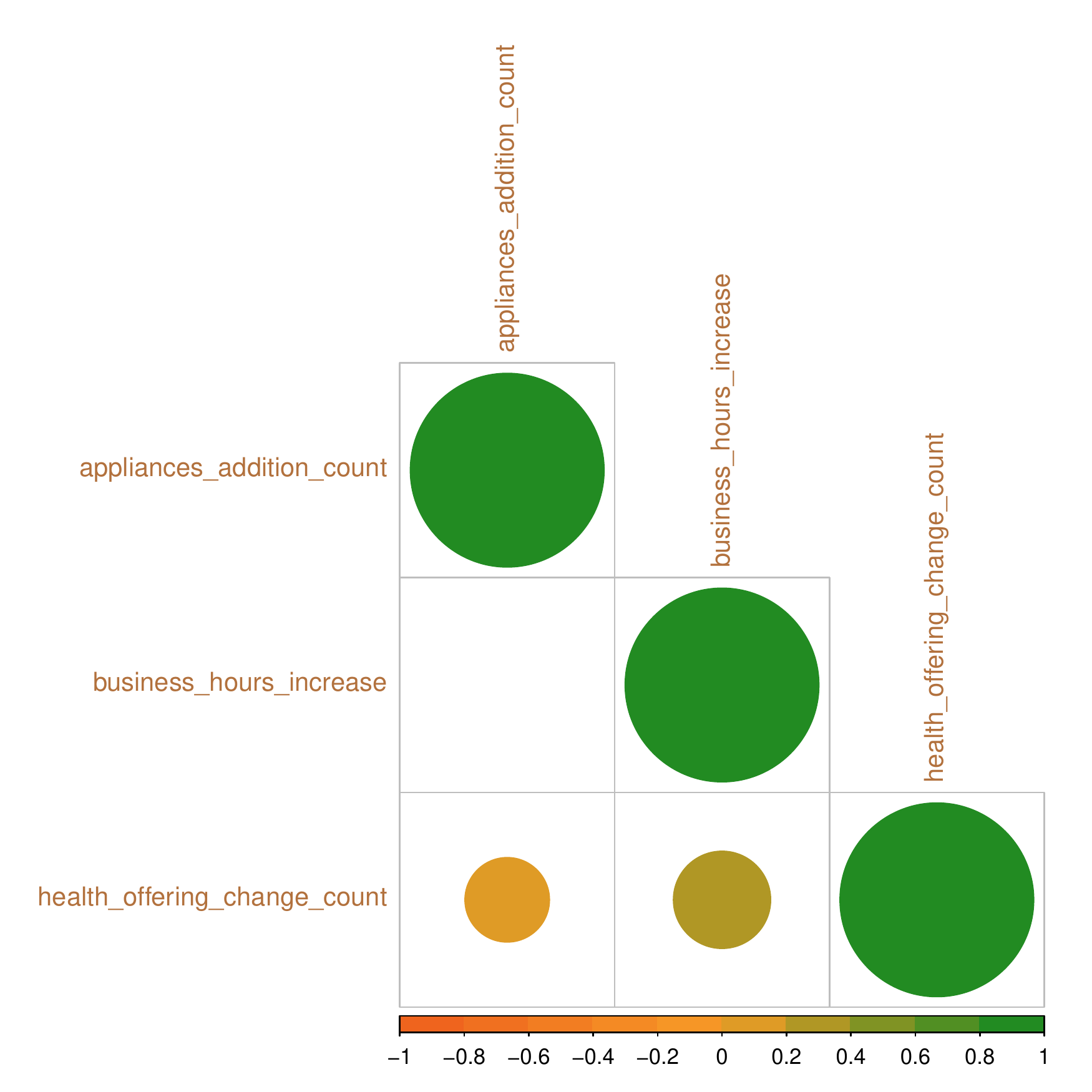}
	\caption{Correlation Matrix of C\&I Post-Survey Responses to 3 Numerical Questions}
	\label{fig:corr-post-ci}
\end{figure}

\begin{landscape}
	\begin{center}
	\begin{longtable}[!ht]{VSTEPlll}
		\caption{Statistical Testing Results for Gender Equality} \label{tab:app:gender} \\
		
		\toprule
		Variable & Sample & Test & Estimate & Predictor & Result & 95\% CI & $p$-value\\
		\midrule
		\endfirsthead

		\multicolumn{8}{c}{\tablename\ \thetable{}, Continued} \\
		\toprule
		Variable & Sample & Test & Estimate & Predictor & Result & 95\% CI & \textit{p}-value\\
		\midrule
		\endhead

		\bottomrule \multicolumn{8}{c}{\appFootnote Table continues on next page.}
		\endfoot

		\multicolumn{8}{c}{\appFootnote}
		\endlastfoot
		
		\multirow[t]{3}{\cwV}{Increase in Schooling for Girls} & \multirow[t]{3}{\cwS}{Post} & \multirow[t]{3}{\cwT}{Linear Regression} & \multirow[t]{3}{\cwE}{Slope of Regression Line} & PV Size (kWp) & 1.824* & (0.425, 3.223) & 0.0140 \\
		&       &        &         & Customer Count & 0.0680 & (0.1181 0.2541) & 0.448\\
		&       &        &         & CAPEX (k USD)  & 0.4134** & (0.1793, 0.6475) & 0.0020\\
		\hline
			
		\multirow[t]{3}{\cwV}{Increase in Schooling for Boys} & \multirow[t]{3}{\cwS}{Post} 
		& \multirow[t]{3}{\cwT}{Linear Regression} & \multirow[t]{3}{\cwE}{Slope of Regression Line} & PV Size (kWp) & 1.764* & (0.273, 3.256)& 0.0235\\
		&       &        &         & Customer Count & 0.0674 & (-0.1252, 0.2601) & 0.467\\
		&       &        &         & CAPEX (k USD)  & 0.4184** & (0.1722, 0.6645) & 0.0027\\
		\hline
			
		\multirow[t]{4}{\cwV}{Water Collection Time} & \multirow[t]{4}{\cwS}{Paired}  
		& Paired $t$-test & Mean Difference & RMG Presence & 0.1483*** & (0.0708, 0.2259) & 0.0002\\
		&        & \multirow[t]{3}{\cwT}{Linear Regression} & \multirow[t]{3}{\cwE}{Slope of Regression Line} & PV Size (kWp) & 2.455* & (0.2576, 4.652)& 0.0320\\
		&        &        &                 & Customer Count & 0.0735 & (-0.2225, 0.3695) & 0.592\\
		&        &        &                 & CAPEX (k USD) & 0.3348** & (0.0279, 0.6418) & 0.0354\\
		\hline
			
		\multirow[t]{4}{\cwV}{Water Collection Responsibility (School-Aged vs. Not)} & \multirow[t]{4}{\cwS}{Paired}  
				& McNemar & Odds Ratio & RMG Presence & 0.2894*** & (0.2104,0.3927) & 2e-16\\
		&       & \multirow[t]{3}{\cwT}{Linear Regression} & \multirow[t]{3}{\cwE}{Slope of Regression Line} & PV Size (kWp) & 4.019 &(-0.8265, 8.892) & 0.0942\\
		&       &        &                 & Customer Count & 0.2279 &(-0.3544, 0.8117) & 0.4035\\
		&       &        &                 & CAPEX (k USD) & 0.0333 & (-0.7445, 0.8111)& 0.9259\\
		\hline
			
		\multirow[t]{4}{\cwV}{Cooking Fuel Collection Time} & \multirow[t]{4}{\cwS}{Paired}
		& Paired $t$-test & Mean Difference & RMG Presence &0.3610*** &(0.2668, 0.4551)& 4.3e-13\\
		&       & \multirow[t]{3}{\cwT}{Linear Regression} &  \multirow[t]{3}{\cwE}{Slope of Regression Line} & PV Size (kWp) & 0.0440 & (-4.2914, 4.3795)&0.9824\\
		&       &        &                 & Customer Count & -0.0134 & (-0.4792, 0.4524)& 0.9503\\
		&       &        &                 & CAPEX (k USD) & 0.2247 & (-0.3544, 0.8037) &0.4076\\
		\pagebreak
			
		Women-led Business Ownership & Paired 
		& McNemar & Odds Ratio &RMG Presence & 0.7957 & (0.5783, 1.092) & 0.164\\
		\hline
		
		\multirow[t]{4}{\cwV}{Employment Opportunity for Women} & \multirow[t]{4}{\cwS}{Post} 
		& Likelihood-Ratio Test & Odds Ratio & Consumption (kWh) & 0.0036 & (-0.0071, 0.0126)& 0.4618\\
		&       & \multirow[t]{3}{\cwT}{Linear Regression} & \multirow[t]{3}{\cwE}{Slope of Regression Line} & PV Size (kWp) & -0.1333 & (-0.8511, 0.5845)& 0.6878\\
		&       &        &          & Customer Count & 0.0697* & (0.0155, 0.1239) & 0.0168 \\
		&       &        &          & CAPEX (k USD) & 0.0334 & (-0.0686, 0.1353) & 0.4825\\
		\bottomrule
	\end{longtable}
	\end{center}
\end{landscape}

\pagebreak

\begin{landscape}
	\begin{center}
	\begin{longtable}[ht]{VSTEPlll}
		\caption{Statistical Testing Results for Productivity} \label{tab:app:productivity} \\
		
		\toprule
		Variable & Sample & Test & Estimate & Predictor & Result & 95\% CI & $p$-value\\
		\midrule
		\endfirsthead

		\multicolumn{8}{c}{\tablename\ \thetable{}, Continued} \\
		\toprule
		Variable & Sample & Test & Estimate & Predictor & Result & 95\% CI & $p$-value \\
		\midrule
		\endhead

		\bottomrule \multicolumn{8}{c}{\appFootnote Table continues on next page.}
		\endfoot

		\multicolumn{8}{c}{\appFootnote}
		\endlastfoot
		
		Light Hours & Paired 
		& Paired \textit{t-}test &Mean Difference &RMG Presence & -3.1211*** & (-3.5087,-2.7336) & 2e-16 \\
		\hline
		
		\multirow[t]{4}{\cwV}{Power Source (Clean vs Unclean)} & \multirow[t]{4}{\cwS}{Paired} 
		& McNemar & Odds Ratio & RMG Presence & 0.1449*** & (0.0954, 0.2132) & 2e-16 \\
		&       & \multirow[t]{3}{\cwT}{Linear Regression} & \multirow[t]{3}{\cwE}{Slope of Regression Line} & PV Size (kWp) & -1.398 & (-6.4602, 3.6642)& 0.5521\\
		&       &        &          & Customer Count &0.4686* & (0.0235,  0.9137) & 0.0409\\
		&       &        &          & CAPEX (k USD)  & 0.2525 & (-0.4390, 0.9440) & 0.4348 \\
		\hline
			
		\multirow[t]{4}{\cwS}{Appliances Count} & \multirow[t]{4}{\cwT}{Paired} 
		& Paired $t$-test & Mean Difference & RMG Presence & -1.5867*** & (-1.7789, -1.3945) & 2e-16 \\
		&       & \multirow[t]{3}{\cwT}{Linear Regression} & \multirow[t]{3}{\cwE}{Slope of Regression Line} & PV Size (kWp) & 0.2424 & (-6.8040, 7.2888)& 0.9404 \\
		&       &        &          & Customer Count & 0.8710** & (0.4270, 1.3149) & 0.0014 \\
		&       &        &          & CAPEX (k USD) & 0.6005 & (-0.2788, 0.1480) & 0.1591 \\
		\hline
	
		\multirow[t]{4}{\cwV}{Phone Charging Travel Distance} & \multirow[t]{4}{\cwS}{Paired} 
		& Wilcoxon & Median & RMG Presence & 0.9999*** & (0.9999, 1.0000) & 2e-16 \\
		&       & \multirow[t]{3}{\cwT}{Linear Regression} & \multirow[t]{3}{\cwE}{Slope of Regression Line} & PV Size (kWp) & -4.410  & (-10.7694, 1.9484) &0.1533\\
		&       &        &          & Customer Count & 0.0901 &(-0.6679, 0.8480) & 0.7965\\
		&       &        &          & CAPEX (k USD) & -0.1692 &(-0.1142, 0.8034) & 0.7064\\
		\hline
	
		\multirow[t]{4}{\cwV}{Water Collection Travel Distance} & \multirow[t]{4}{\cwS}{Paired} 
		& Wilcoxon & Median & RMG Presence & 6.5e-6* & (-5.2e-5, 6.7e-7) & 0.0429 \\
		&       & \multirow[t]{3}{\cwT}{Linear Regression} & \multirow[t]{3}{\cwE}{Slope of Regression Line} & PV Size (kWp) & 0.8506 & (-2.0535, 3.7547) & 0.5287 \\
		&       &        &          & Customer Count & 0.0472 & (-0.2697, 0.3641) & 0.7470\\
		&       &        &          & CAPEX (k USD) & 0.2094 & (-0.1738, 0.5925) &0.2513 \\
		\hline
	
		Clinic Travel Distance & Paired 
		& Wilcoxon & Median & RMG Presence & -3.6e-5 & (-2.8e-5, 6.1e-7) & 0.9765\\
		\hline
			
		\multirow[t]{3}{\cwV}{Academic Performance (Households)} & \multirow[t]{3}{\cwS}{Post} 
		& \multirow[t]{3}{\cwT}{Linear Regression} & \multirow[t]{3}{\cwE}{Slope of Regression Line} & PV Size (kWp) & 1.4668 & (-0.1765, 3.1100) & 0.0765\\
		&       &        &          & Customer Count & 0.1874* & (0.0139, 0.3609) & 0.0361\\
		&       &        &          & CAPEX (k USD) & 0.3049* & (0.2175, 0.5880) & 0.0367 \\
		\pagebreak
			
		\multirow[t]{3}{\cwV}{Academic Performance (Schools)} & \multirow[t]{3}{\cwS}{Post} 
		& \multirow[t]{3}{\cwT}{Linear Regression} & \multirow[t]{3}{\cwE}{Slope of Regression Line} & PV Size (kWp) & 0.2344** & (0.0803, 0.3885) &0.0074\\
		&       &        &          & Customer Count & -0.0089 & (-0.0355, 0.0178)& 0.4712\\
		&       &        &          & CAPEX (k USD) & 0.0201  & (-0.017, 0.0572) & 0.2514\\
		\hline
			
		\multirow[t]{3}{\cwV}{Hours of Operations} & \multirow[t]{3}{\cwS}{Post} 
		& \multirow[t]{3}{\cwT}{Linear Regression} & \multirow[t]{3}{\cwE}{Slope of Regression Line} & PV Size (kWp) & 1.0059 & (-0.0224, 2.0343) & 0.0545\\
		&       &        &          & Customer Count & 0.0872 & (-0.0439, 0.2183) & 0.1758\\
		&       &        &          & CAPEX (k USD) & 0.0568 & (-0.1396, 0.2533) & 0.5449\\
		\hline
			
		\multirow[t]{3}{\cwV}{Employment Opportunity For All} & \multirow[t]{3}{\cwS}{Post} 
		& \multirow[t]{3}{\cwT}{Linear Regression} & \multirow[t]{3}{\cwE}{Slope of Regression Line} & PV Size (kWp) & 0.0348 & (-0.3229, 0.3925)& 0.8344\\
		&       &        &          & Customer Count & 0.0382* & (0.0090, 0.0674) & 0.0149\\
		&       &        &          & CAPEX (k USD) & 0.0039 & (-0.0511, 0.0589) & 0.8783\\
		\hline
			
		\multirow[t]{3}{\cwV}{Change in Products/Services} & \multirow[t]{4}{\cwS}{Post} 
		& Likelihood-Ratio Test & Odds Ratio & Consumption (kWh) & 0.0037 & (-0.0045, 0.0133)& 0.3808\\
		&       & \multirow[t]{3}{\cwT}{Linear Regression} & \multirow[t]{3}{\cwE}{Slope of Regression Line} & PV Size (kWp) & 0.8667** &(0.3663, 1.3671) & 0.0023\\
		&       &        &          & Customer Count & 0.0376 & (-0.0435, 0.1187) & 0.3374\\
		&       &        &          & CAPEX (k USD) & 0.0743 & (-0.0369, 0.1856) & 0.1737\\
		
		\bottomrule
	\end{longtable}
	\end{center}
\end{landscape}

\pagebreak

\begin{landscape}
	\begin{center}
	\begin{longtable}[ht]{VSTEPlll}
		\caption{Statistical Testing Results for Health} \label{tab:app:health} \\
		
		\toprule
		Variable & Sample & Test & Estimate & Predictor & Result & 95\% CI & $p$-value\\
		\midrule
		\endfirsthead

		\multicolumn{8}{c}{\tablename\ \thetable{}, Continued} \\
		\toprule
		Variable & Sample & Test & Estimate & Predictor & Result & 95\% CI & $p$-value \\
		\midrule
		\endhead

		\bottomrule \multicolumn{8}{c}{\appFootnote Table continues on next page.}
		\endfoot

		\multicolumn{8}{c}{\appFootnote}
		\endlastfoot
		
		\multirow[t]{4}{\cwV}{Access to Clean Drinking Water}  & \multirow[t]{4}{\cwS}{Paired} 
		& McNemar & Odds Ratio & RMG Presence & 1.8689*** & (1.3577, 2.5941) & 0.0001 \\
		&       & \multirow[t]{3}{\cwT}{Linear Regression} & \multirow[t]{3}{\cwE}{Slope of Regression Line} & PV Size (kWp) & -1.364 & (-6.0004, 3.2721) &0.5269 \\
		&       &        &          & Customer Count & -0.1371 & (-0.6367, 0.3624)& 0.5545 \\
		&       &        &          & CAPEX (k USD) & -0.1735 & (-0.8175, 0.4706) & 0.5618 \\
		\hline
			
		\multirow[t]{4}{\cwV}{Kerosene Lamps} & \multirow[t]{4}{\cwS}{Paired} 
		& Paired $t$-test & Mean Difference & RMG Presence & 0.3627*** & (0.1658, 0.5597) & 0.0004 \\
		&       & \multirow[t]{3}{\cwT}{Linear Regression} & \multirow[t]{3}{\cwE}{Slope of Regression Line} & PV Size (kWp) & -2.148 & (-6.2876, 1.9919) &0.2745\\
		&       &        &          & Customer Count & 0.0951 & (-0.3739, 0.5639) & 0.6612\\
		&       &        &          & CAPEX (k USD) & -0.3178 &(-0.8855, 0.2499) & 0.2407\\
		\hline
			
		\multirow[t]{3}{\cwV}{Kerosene Lamp Usage} & \multirow[t]{3}{\cwS}{Paired}       
		& \multirow[t]{3}{\cwT}{Linear Regression} & \multirow[t]{3}{\cwE}{Slope of Regression Line} & PV Size (kWp) & -3.6820 &(-8.4022, 1.0379) &0.1171\\
		&       &        &          & Customer Count &0.7038**  & (0.2907, 1.1170) & 0.0025 \\
		&       &        &          & CAPEX (k USD) & -0.2940 & (-1.1234, 0.5355)&0.4598 \\
		\hline
		
		\multirow[t]{3}{\cwV}{New Health Services} & \multirow[t]{3}{\cwS}{Paired}       
		& \multirow[t]{3}{\cwT}{Linear Regression} & \multirow[t]{3}{\cwE}{Slope of Regression Line} & PV Size (kWp) & 0.0262 & (-0.1488, 0.2012) & 0.7335 \\
		&       &        &          & Customer Count & 0.0062 & (-0.0125, 0.0249) & 0.4585\\
		&       &        &          & CAPEX (k USD) & -0.0016 & (-0.0318, 0.0285) & 0.9007\\
		\hline
		
		Clinic Access To Electricity  & Paired 
		& McNemar & Odds Ratio &RMG Presence &117*** & (32.03, 971.8) & 2e-16\\
		\hline
		
		Clinic Access To Refrigeration  & Paired 
		& McNemar & Odds Ratio &RMG Presence  & 11.895*** & ( 7.4418, 20.126)& 2e-16 \\
		\hline
		
		\multirow[t]{4}{\cwV}{Better Access to Healthcare} & \multirow[t]{4}{\cwS}{Post}
		& Likelihood-Ratio Test & Log Odds & Consumption (kWh) & -0.0174 & (-0.0513, 0.0316) & 0.4269 \\
		& & \multirow[t]{3}{\cwT}{Linear Regression} & \multirow[t]{3}{\cwE}{Slope of Regression Line} & PV Size (kWp) & -0.949 & (-5.0190, 3.1209) &0.6264\\
		&       &        &          & Customer Count &0.7576*** & (0.5821, 0.9331) & 1.48e-7\\
		&       &        &          & CAPEX (k USD) & 0.1364 & (-0.5194, 0.7922) & 0.6623\\

		\multirow[t]{3}{\cwV}{Self-Reported Improvement} & \multirow[t]{3}{\cwS}{Post} 
		& Likelihood-Ratio Test & Log Odds & Consumption (kWh) & -0.0140 & (-0.0379, 0.0089)& 0.2299\\
		& & \multirow[t]{3}{\cwT}{Linear Regression} & \multirow[t]{3}{\cwE}{Slope of Regression Line} & PV Size (kWp) & 0.7911 & (-1.8564, 3.4386) &0.5338\\
		&       &        &          & Customer Count &0.4694*** & (0.3253, 0.6135) & 4.72e-6\\
		&       &        &          & CAPEX (k USD) & 0.2127 & (-0.2632, 0.6886) & 0.3540\\
	
		\bottomrule
	\end{longtable}
	\end{center}
\end{landscape}

\pagebreak

\begin{landscape}
	\begin{center}
	\begin{longtable}[ht]{VSTEPlll}
		\caption{Statistical Testing Results for Safety} \label{tab:app:safety} \\
		
		\toprule
		Variable & Sample & Test & Estimate & Predictor & Result & 95\% CI & $p$-value\\
		\midrule
		\endfirsthead

		\multicolumn{8}{c}{\tablename\ \thetable{}, Continued} \\
		\toprule
		Variable & Sample & Test & Estimate & Predictor & Result & 95\% CI & $p$-value \\
		\midrule
		\endhead

		\bottomrule \multicolumn{8}{c}{\appFootnote Table continues on next page.}
		\endfoot

		\multicolumn{8}{c}{\appFootnote}
		\endlastfoot
		
		\multirow[t]{4}{\cwV}{Feeling Unsafe Due to Unsafe Travel} & \multirow[t]{4}{\cwS}{Paired} 
		& McNemar & Odds Ratio & RMG Presence & 0.1759*** & (0.1020, 0.2882) & 2.7e-16\\
		&       & \multirow[t]{3}{\cwT}{Linear Regression} & \multirow[t]{3}{\cwE}{Slope of Regression Line} & PV Size (kWp) & 1.0250 & (-1.7876, 3.8377) &0.4357\\
		&       &        &          & Customer Count & -0.0609 & (-0.3700, 0.2482) & 0.6699\\
		&       &        &          & CAPEX (k USD) & 0.1351 & (-0.2555, 0.5257) &0.4587\\
		\hline
		
		\multirow[t]{4}{\cwV}{Feeling Unsafe Due to No Community Lighting} & \multirow[t]{4}{\cwS}{Paired} 
		& McNemar & Odds Ratio & RMG Presence &0.4094*** &(0.2988, 0.5550)&1.1e-9\\ 
		&       & \multirow[t]{3}{\cwT}{Linear Regression} & \multirow[t]{3}{\cwE}{Slope of Regression Line} & PV Size (kWp) &-3.145&(-7.5491, 1.2582) & 0.1426 \\
		&       &        &          & Customer Count &0.3992  &(-0.0497, 0.8482) & 0.0757 \\
		&       &        &          & CAPEX (k USD) & -0.1154 &(-0.7931, 0.5623) & 0.7123 \\
		\hline
	
		\multirow[t]{4}{\cwV}{Feeling Unsafe Due to Potential Theft}  & \multirow[t]{4}{\cwS}{Paired}
		& McNemar & Odds Ratio &RMG Presence & 5.925*** & (4.2245 8.5047) & 2e-16 \\
		&       & \multirow[t]{3}{\cwT}{Linear Regression} & \multirow[t]{3}{\cwE}{Slope of Regression Line} & PV Size (kWp) & 0.9278 &(-0.7589, 2.6144) &0.2484\\
		&       &        &          & Customer Count & -0.0562 &(-0.2465, 0.1341) & 0.5252\\
		&       &        &          & CAPEX (k USD) & 0.1628 &(-0.0599, 0.3854) & 0.1344\\
		\hline
		
		\multirow[t]{4}{\cwV}{Presence of Home Exterior Lights} & \multirow[t]{4}{\cwS}{Paired} 
		& McNemar & Odds Ratio &RMG Presence & 4.1395*** &(2.9526, 5.9187) & 2e-16 \\
		&       & \multirow[t]{3}{\cwT}{Linear Regression} & \multirow[t]{3}{\cwE}{Slope of Regression Line} & PV Size (kWp) &-1.141  &(-5.8565, 3.5753) &0.6018\\
		&       &        &          & Customer Count & 0.3410 &(-0.1134, 0.7955) & 0.1255\\
		&       &        &          & CAPEX (k USD) & 0.1380 &(-0.5171, 0.7932) & 0.6489\\
	
		\bottomrule
	\end{longtable}
	\end{center}
\end{landscape}

\begin{landscape}
	\begin{center}
	\begin{longtable}[ht]{VSTEPlll}
		\caption{Statistical Testing Results for Economic Activity} \label{tab:app:econ} \\
		
		\toprule
		Variable & Sample & Test & Estimate & Predictor & Result & 95\% CI & $p$-value\\
		\midrule
		\endfirsthead

		\multicolumn{8}{c}{\tablename\ \thetable{}, Continued} \\
		\toprule
		Variable & Sample & Test & Estimate & Predictor & Result & 95\% CI & $p$-value \\
		\midrule
		\endhead

		\bottomrule \multicolumn{8}{c}{\appFootnote Table continues on next page.}
		\endfoot

		\multicolumn{8}{c}{\appFootnote}
		\endlastfoot
		
		\multirow[t]{4}{\cwV}{Average Household Income - Nigeria} & \multirow[t]{4}{\cwS}{Paired} 
		& Paired $t$-test & Mean Difference & RMG Presence & 24080*** & (19583, 28574) & 2e-16 \\
		&       & \multirow[t]{3}{\cwT}{Linear Regression} & \multirow[t]{3}{\cwE}{Slope of Regression Line} & PV Size (kWp) &0.3541 &(-5.4645, 6.1727) &0.8179 \\
		&       &        &          & Customer Count & 0.0896 & (-0.3758, 0.5550) & 0.4946\\
		&       &        &          & CAPEX (k USD) & -0.2976 & (-8.4745, 7.8792) & 0.7242 \\
		\hline
		
		\multirow[t]{4}{\cwV}{Average Household Income - Kenya} & \multirow[t]{4}{\cwS}{Paired} 
		& Paired $t$-test & Mean Difference & RMG Presence & -13037*** & (-11375, -14698) & 2e-16 \\
		&       & \multirow[t]{3}{\cwT}{Linear Regression} & \multirow[t]{3}{\cwE}{Slope of Regression Line} & PV Size (kWp) & 1.0640 &(-4.4101, 6.5380) &0.6511\\
		&       &        &          & Customer Count & 0.4714** & (0.1781, 0.7648) & 0.0077\\
		&       &        &          & CAPEX (k USD) & 0.0867 & (-0.6213, 0.7946) &0.7747\\
		\hline

		\multirow[t]{3}{\cwV}{Household Income Signed Change} & \multirow[t]{3}{\cwS}{Post}
		& \multirow[t]{3}{\cwT}{Linear Regression} & \multirow[t]{3}{\cwE}{Slope of Regression Line} & PV Size (kWp) & 0.7668 & (-0.7463, 2.2798) & 0.2971\\
		&       &        &          & Customer Count & 0.1416 &(-0.0130, 0.2963) &0.0698 \\
		&       &        &          & CAPEX (k USD) &0.1492  &(-0.0065, 0.3050) & 0.0590\\
		\hline
		
		\multirow[t]{3}{\cwV}{Business Earnings Signed Change} & \multirow[t]{3}{\cwS}{Post} 
		& \multirow[t]{3}{\cwT}{Linear Regression} & \multirow[t]{3}{\cwE}{Slope of Regression Line} & PV Size (kWp) &1.1001* &(0.1734, 2.0269) & 0.0233\\
		&       &        &          & Customer Count & 0.0904 &(-0.0325, 0.2133) & 0.1370\\
		&       &        &          & CAPEX (k USD) & 0.0894 &(-0.0928, 0.2717) & 0.3105\\
		\hline
			
		\multirow[t]{4}{\cwV}{Business Ownership} & \multirow[t]{4}{\cwS}{Paired} 
		& McNemar & Odds Ratio &RMG Presence & 6.1111*** &(2.9962, 14.0661) & 3.5e-5 \\
		&       & \multirow[t]{3}{\cwT}{Linear Regression} & \multirow[t]{3}{\cwE}{Slope of Regression Line} & PV Size (kWp) & -0.4382 &(-5.8828, 5.0065) &0.8613\\
		&       &        &          & Customer Count & 0.1741 & (-0.3990, 0.7471) & 0.5139\\
		&       &        &          & CAPEX (k USD) & -0.4673 &(-1.14676, 0.2122) & 0.1564\\
		\hline
	
		\multirow[t]{4}{\cwV}{Phone Charging Cost - Nigeria} & \multirow[t]{4}{\cwS}{Paired} 
		& Sign & Median &RMG Presence &-4*** & (-5, -4) & 9e-52\\
		&       & \multirow[t]{3}{\cwT}{Linear Regression} & \multirow[t]{3}{\cwE}{Slope of Regression Line} & PV Size (kWp) & 7.729& (-0.0621, 15.5200) & 0.0507\\
		&       &        &          & Customer Count & 0.7266* & (0.2671, 1.1861) & 0.0209\\
		&       &        &          & CAPEX (k USD) &0.7277 & (-4.4249, 5.8804) & 0.3236\\
		\pagebreak
		
		\multirow[t]{4}{\cwV}{Phone Charging Cost - Kenya} & \multirow[t]{4}{\cwS}{Paired} 
		& Sign & Median &RMG Presence &-1** & (-1, -1) & 0.0022\\
		&      & \multirow[t]{3}{\cwT}{Linear Regression} & \multirow[t]{3}{\cwE}{Slope of Regression Line} & PV Size (kWp) &-4.4173 &(-11.3411, 2.5066) & 0.3741\\
		&       &        &          & Customer Count &-0.2931 & (-1.0558, 0.4696) & 0.3834\\
		&       &        &          & CAPEX (k USD) &-0.6168 & (-1.4672, 0.2335) & 0.1263\\
		\hline
		
		\multirow[t]{4}{\cwV}{Appliances Charging Cost - Nigeria} & \multirow[t]{4}{\cwS}{Paired} 
		& Sign & Median &RMG Presence &-3*** & (-3, -3) & 1e-50\\
		&       & \multirow[t]{3}{\cwT}{Linear Regression} & \multirow[t]{3}{\cwE}{Slope of Regression Line} & PV Size (kWp) &7.4115 &(-1.9588, 16.7818) & 0.0765\\
		&       &        &          & Customer Count &0.7266 & (0.4684, 0.9848) & 0.0067\\
		&       &        &          & CAPEX (k USD) &0.7450 & (-9.764, 11.254) & 0.5332 \\
		\hline
		
		\multirow[t]{4}{\cwV}{Appliances Charging Cost - Kenya} & \multirow[t]{4}{\cwS}{Paired} 
		& Sign & Median &RMG Presence &-1*** & (-1,-1) & 2.60e-6\\
		&       & \multirow[t]{3}{\cwT}{Linear Regression} & \multirow[t]{3}{\cwE}{Slope of Regression Line} & PV Size (kWp) &-4.0440 &(-11.5460, 3.4581) & 0.2353\\
		&       &        &          & Customer Count &-0.2322 & (-1.0476, 0.5833) & 0.5121\\
		&       &        &          & CAPEX (k USD) &-0.6206 & (-1.5169, 0.2757) & 0.1412\\
		\hline

		\multirow[t]{4}{\cwV}{Cooking Energy Cost - Nigeria} & \multirow[t]{4}{\cwS}{Paired} 
		& Wilcoxon & Median &RMG Presence &1.49*** & (0.99, 1.5) & 6e-12\\
		&       & \multirow[t]{3}{\cwT}{Linear Regression} & \multirow[t]{3}{\cwE}{Slope of Regression Line} & PV Size (kWp) & 4.560 & (-23.6620, 32.7829) & 0.5588\\
		&       &        &          & Customer Count &0.6825 & (-1.293, 2.658) & 0.2755\\
		&       &        &          & CAPEX (k USD) & -0.3187 & (-43.6965, 43.059) & 0.9407\\
		\hline
		
		\multirow[t]{4}{\cwV}{Cooking Energy Cost - Kenya} & \multirow[t]{4}{\cwS}{Paired} 
		& Wilcoxon & Median &RMG Presence &-4.89e-5** & (-0.99, -1.91e-5) & 0.001442\\
		&       & \multirow[t]{3}{\cwT}{Linear Regression} & \multirow[t]{3}{\cwE}{Slope of Regression Line} & PV Size (kWp) &0.2391 &(-2.2511, 2.7294
) &0.8221\\
		&       &        &          & Customer Count &0.0074 & (-0.2414, 0.2562) &0.9442 \\
		&       &        &          & CAPEX (k USD) &0.0295 & (-0.2891, 0.3481) &0.8285 \\
		\hline

		\multirow[t]{4}{\cwV}{Kerosene Lamps Cost - Nigeria} & \multirow[t]{4}{\cwS}{Paired}  
		& Sign & Median &RMG Presence &-3*** & (-4, -3) & 1e-20\\
		&       & \multirow[t]{3}{\cwT}{Linear Regression} & \multirow[t]{3}{\cwE}{Slope of Regression Line} & PV Size (kWp) & 6.0134 &(-2.674, 14.701) & 0.0967\\
		&       &        &          & Customer Count & 0.6042** & (0.4334, 0.7749) & 4e-3 \\
		&       &        &          & CAPEX (k USD) &0.734 & (-11.1702, 12.6383) & 0.5769\\
		\pagebreak
		
		\multirow[t]{4}{\cwV}{Kerosene Lamps Cost - Kenya} & \multirow[t]{4}{\cwS}{Paired} 
		& Sign & Median &RMG Presence &-1*** & (-1, -1) & 5e-12\\
		&       & \multirow[t]{3}{\cwT}{Linear Regression} & \multirow[t]{3}{\cwE}{Slope of Regression Line} & PV Size (kWp) &-3.1656 & (-9.5667, 3.2356) &  0.2718\\
		&       &        &          & Customer Count &-0.1546 & (-0.8480, 0.5388) &0.6051\\
		&       &        &          & CAPEX (k USD) & -0.6074 & (-1.2898, 0.0750) & 0.0723\\
		\hline

		\multirow[t]{4}{\cwV}{Water Cost - Nigeria} & \multirow[t]{4}{\cwS}{Paired} 
		& Sign & Median &RMG Presence &-2*** & (-2, -2) & 3e-4 \\
		&       & \multirow[t]{3}{\cwT}{Linear Regression} & \multirow[t]{3}{\cwE}{Slope of Regression Line} & PV Size (kWp) &2.2100 &(-0.6804, 5.1004) & 0.0812\\
		&       &        &          & Customer Count &0.2098* & (0.0157, 0.4039) &  0.0433\\
		&       &        &          & CAPEX (k USD) &0.1136 & (-1.138, 1.365) & 0.4547 \\
		\hline
		
		\multirow[t]{4}{\cwV}{Water Cost - Kenya} & \multirow[t]{4}{\cwS}{Paired} 
		& Sign & Median &RMG Presence &-0.5 & (-1.0, -1.7e-5) & 0.0062\\
		&       & \multirow[t]{3}{\cwT}{Linear Regression} & \multirow[t]{3}{\cwE}{Slope of Regression Line} & PV Size (kWp) &0.3696 & (-3.8569, 4.5962) &0.8376\\
		&       &        &          & Customer Count &-0.1981 & (-0.5710, 0.1748) & 0.2414\\
		&       &        &          & CAPEX (k USD) &0.1584 & (-0.3607, 0.6774) &0.4835\\
		\hline

		\bottomrule
	\end{longtable}
	\end{center}
\end{landscape}
\end{document}